\documentclass[journal]{IEEEtran}

\usepackage{color}
\usepackage{xcolor}
\usepackage{cite}
\usepackage{graphicx}
\usepackage{amsmath}
\usepackage{amssymb}
\usepackage{algorithm}
\usepackage{algpseudocode}
\usepackage{multirow}
\usepackage{tikz}
\usepackage{hyperref}
\usepackage{geometry}

\IEEEoverridecommandlockouts

\newcommand\copyrighttext{%
\footnotesize \textcopyright 2021 IEEE. Personal use of this material is permitted. Permission from IEEE must be obtained for all other uses, in any current or future media, including reprinting/republishing this material for advertising or promotional purposes, creating new collective works, for resale or redistribution to servers or lists, or reuse of any copyrighted component of this work in other works. DOI: \href{https://ieeexplore.ieee.org/document/9578939}{10.1109/TCSI.2021.3119285}}
\newcommand\copyrightnotice{%
\begin{tikzpicture}[remember picture,overlay]
\node[anchor=south,yshift=10pt] at (current page.south) {\fbox{\parbox{\dimexpr\textwidth-\fboxsep-\fboxrule\relax}{\copyrighttext}}};
\end{tikzpicture}%
}

\begin{document}

\title{Adaptive Filtering with Reduced Computational Complexity Using SOPOT Arithmetic}

\author{Luiz~Felipe~da~S.~Coelho,~\IEEEmembership{Student Member,~IEEE,}
        Lisandro~Lovisolo,~\IEEEmembership{Senior Member,~IEEE,}
        and~Michel~P.~Tcheou,~\IEEEmembership{Member,~IEEE}
\thanks{L. F. S. Coelho is with the Department of Electronic and Computer Engineering, Universidade Federal do Rio de Janeiro (UFRJ), RJ, 21941-972, Brazil. L. Lovisolo and M. P. Tcheou are with the Universidade do Estado do Rio de Janeiro (UERJ), RJ, 21945-970, Brazil. Email: luizfelipe.coelho@smt.ufrj.br, \{lisandro, mtcheou\}@uerj.br}
\thanks{This work was partially supported by Conselho Nacional de Desenvolvimento Científico e Tecnológico (CNPq) and by Fundação Carlos Chagas Filho de Amparo à Pesquisa do Estado do Rio de Janeiro (FAPERJ).}
}

% \markboth{IEEE Transactions on Circuits and Systems I: Regular Papers,~Vol.~, No.~, month~year}%
% {Shell \MakeLowercase{\textit{et al.}}: Bare Demo of IEEEtran.cls for IEEE Journals}

\newgeometry{top=.7in,bottom=0.9in,right=0.7in,left=.7in}
\maketitle
\copyrightnotice

\begin{abstract}
Implementing \textit{finite impulse response} (FIR) adaptive filters by employing the \textit{sums of signed-powers-of-two} (SOPOT) arithmetic may lead to simpler hardware and consequently reduced power consumption. In this paper, one evaluates the effects of SOPOT arithmetic on the adaptive filter's recursion algorithms. The filters' coefficients and algorithms' underlying variables are fully operated using SOPOT arithmetic in the whole iterative process. More specifically, one evaluates convergence rate, numerical stability, and accuracy since using few \textit{signed-powers-of-two} (SPT) terms propagates numerical errors during the adaptive cycle that may impair the algorithm behavior. The SOPOT approximations are obtained through the technique known as \textit{Matching Pursuits with Generalized Bit-Plane} (MPGBP) algorithm, with notable cost-performance trade-off and low computational complexity. Results are provided for the \textit{Least-Mean-Squares} (LMS), the \textit{Normalized Least-Mean-Squares} (NLMS) and the \textit{Recursive-Least-Squares} (RLS) algorithms, considering adaptive filters employed for system identification and change detection. 
\end{abstract}


\IEEEpeerreviewmaketitle

\section{Introduction}

\IEEEPARstart{T}{he} urge for more efficient resource management drives many research fields of engineering. This work explores a way to reduce power consumption in adaptive filtering. This is relevant for the development of more power-efficient devices for \textit{smart cities} scenarios~\cite{yaqoob2017enabling} and IoT (Internet-of-Things) applications for healthcare, smart building, and automotive purposes~\cite{henkel2017ultra,samie2016iot}. 

In particular, representing filter coefficients as \textit{sums of signed powers-of-two} (SOPOT) instead of using high-precision floating-point coefficients allows efficient hardware implementation~\cite{samueli1989improved,evans1994efficient,macleod2005multiplierless,lim1999signed,yli2001systematic}. This is due to the multiplier-less nature of SOPOT arithmetic, which allows filter coefficients to be generated from a combination of shifts, additions, and subtractions of \textit{signed powers-of-two} (SPT) terms, i.e., no multiplier is required whereas still letting to program/update the filter coefficients on the fly~\cite{yeung2002multiplier}. Moreover, SOPOT representation has been applied in the implementation of several finite impulse response (FIR) filters~\cite{samueli1989improved,evans1994efficient,macleod2005multiplierless,lim1999signed,yli2001systematic,da2014fir}. In addition to the aforementioned examples, a 2D wavelet transform was designed using SPT in~\cite{iwahashi2016optimum}. In this work, we consider using SPT for implementing the \textit{Least Mean Squares} (LMS), the \textit{Normalized Least Mean Squares} (NLMS) and the \textit{Recursive Least Squares} (RLS) with low complexity (a proxy for energy efficiency) FIR adaptive filters.

Work on energy-efficient FIR adaptive filter approximations has begun around using many different strategies. Recently, in~\cite{jiang2018high}, approximately distributed arithmetic circuits were employed for fixed-point finite impulse response adaptive filter design. The so-called Delayed LMS (DLMS), presented in~\cite{meher2013critical}, uses registers between the FIR and the learning part, reducing some demands. In~\cite{di2021novel}, authors represent the error signal using sign-modulus representation to reduce the adaptive feedback switching activity in DLMS. An LMS-based algorithm using SPT terms was proposed in~\cite{li1994multiplierless}. In~\cite{choudhary2012spt}, some hardware considerations for SPT-based sign-LMS algorithms were discussed. One sees that a long period has passed without many advances in this matter.  This paper generalizes previous initiatives using a common design procedure for SOPOT implementation in adaptive filters. We employ the \textit{Matching Pursuits with Generalized Bit-Planes} (MPGBP)~\cite{da2014fir} algorithm not only for the SOPOT approximation of coefficients originated from adaptive filtering but also for the auxiliary variables that take part in the iterative numerical process for the design of the SPT-based adaptive filter. The MPGBP algorithm has the capability of limiting the amount of SPT terms employed to approximate a vector and also the bit-plane depth of its numeric representation, thus presenting advantages when dealing with low-complexity designs. A comparison between different methods for the acquisition of multiplier-less non-uniform channel filters is presented in~\cite{baboji2019optimal}, and among them, the MPGBP algorithm has resulted in designs of smaller complexity using fewer resources. The SOPOT representation is effective for the design of multiplier-less fixed FIR filters~\cite{da2014fir}, and for RLS adaptive filters~\cite{coelho2020proposal}.

This paper unveils the effects of the SOPOT approximations on adaptive filtering, leading to the so-called SPT-based adaptive filters. This work broadens the framework of SPT-based adaptive filtering proposed in~\cite{coelho2020proposal} by extending it to LMS and NLMS algorithms and providing deeper and more comprehensive performance analysis. It also includes the change detection scenario, which is applicable in spectrum sensing, more specifically in spectrum hole detection~\cite{haykin2005cognitive,e2014spectrum}. Filter coefficients evolve/adapt using SOPOT arithmetic. The adaptive filter is designed using SOPOT arithmetic, differently from the approach in~\cite{da2014fir} where approximation takes place after filter design. This might present numerical instability due to the inherent finite precision of the SOPOT representation. Consequently, this issue is investigated in this work. In principle, the fewer SPT terms are employed (i.e., the less dense are the SOPOT approximations), the more coarsely are the value approximations, and the larger is the expected impairment. Thus, the quantization/approximation effects are assessed using the three well-known adaptive filters, the LMS, NLMS, and RLS. The resulting adaptive filters fully operate in SOPOT arithmetic and are referred to as \textit{Signed Power-of-Two Least Mean Squares} (SPT-LMS), \textit{Signed Power-of-Two Normalized Least Mean Squares} (SPT-NLMS), and \textit{Signed Power-of-Two Recursive Least Squares} (SPT-RLS). Results for system identification and change detection applications are provided and are compared with the high precision numeric representation and complexity equivalent fixed-point versions.

This work is organized as follows. Section~\ref{sec:adptFilt} brings the adaptive filters used. Section~\ref{sec:SOPOTApprox} discusses the SOPOT numeric representation and the MPGBP algorithm used for obtaining SOPOT approximations, as well as some preliminary results on the impact of these approximations upon adaptive filters performance. In Section~\ref{sec:spt_based}, the SPT-based adaptive filters are presented along with their implementation aspects, issues, and proposed solutions. In Section~\ref{sec:results}, simulation results are presented in two different applications: system identification and change detection. Section~\ref{sec:conclusion} presents the concluding remarks.

\section{The Adaptive Filters} \label{sec:adptFilt}

\begin{figure}[hbt!]
    \centering
    \includegraphics[width=\linewidth]{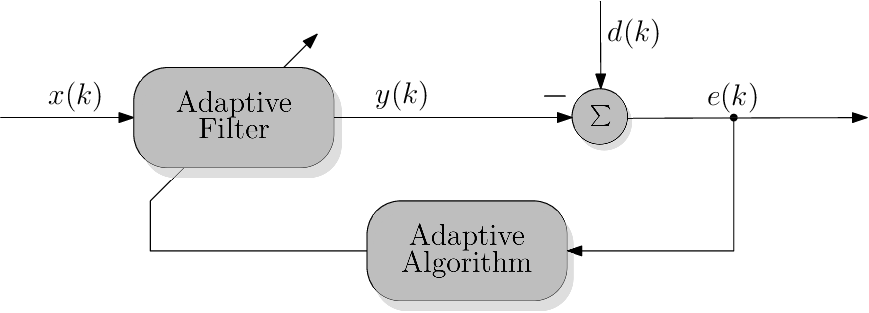}
    \caption{Block diagram for a general adaptive filter, where the filter coefficients are updated using an adaptive algorithm based on the error signal.}
    \label{fig:adaptive_filter}
\end{figure}

Digital filters are commonly represented by a vector containing the $N$ coefficients for a $N-1$ order filter. Filters may have fixed coefficients, but some applications require filters whose coefficients may change, i.e., adapt. These are called adaptive filters. A diagram for adaptive filtering is presented in Figure~\ref{fig:adaptive_filter}, where $k$ indexes the signal samples. An input signal $x(k)$ is filtered, resulting in the output signal $y(k)$. The output signal is then subtracted from the desired signal $d(k)$ resulting in the error signal $e(k)$. An adaptive algorithm governs the update process of the filter coefficients to minimize the error power.

There are many well-known adaptive algorithms~\cite{diniz1997adaptive}. In this work, we consider the LMS, NLMS, and RLS adaptive filters, which are briefly presented in the following.

\subsection{LMS Algorithm}

The LMS is one of the most used algorithms for adaptive filtering. It reduces the \textit{mean squared error} (MSE = $E \left[ |e(k)|^2 \right]$) at the filter output as $k$ increases by computing a simplification of the gradient vector derived using an appropriate change on the objective function~\cite{diniz1997adaptive}. For the LMS, 
\begin{align}
    y(k) &= {\mathbf x}^T(k) {\mathbf w}(k), \label{eq:lms_output} \\ 
    e(k) &= d(k) - y(k), \\
    {\mathbf w}(k+1) &= {\mathbf w}(k) + 2 \mu e(k) {\mathbf x}(k), \label{eq:lms_update}
\end{align}
where ${\mathbf w}(k)$ is an $N$-length column-vector containing the adaptive filter coefficients for the $k$-th iteration, ${\mathbf x}(k)~=~[x(k)~x(k-1)~\cdots~x(k-N+1)]^T$ is the tapped delay line for the input signal in the $k$-th iteration also represented as an $N$-length column vector. Equation (\ref{eq:lms_update}) is the update rule and $\mu$ is the step-size, which is used to adjust the convergence speed and stability of the algorithm and $\cdot^T$ stands for the transpose operator.

\subsection{NLMS Algorithm}

Aiming primarily at faster convergence in the LMS algorithm a variable step-size is employed, this is achieved with a normalization of the tapped delay line in Equation (\ref{eq:lms_update}). The resulting update rule for the NLMS algorithm is
\begin{align}\label{eq:nlms_update}
    {\mathbf w}(k+1) &= {\mathbf w}(k) + \frac{\mu_n e(k) {\mathbf x}(k)}{\gamma + {\mathbf x}^T(k) {\mathbf x}(k)},
\end{align}
where $\mu_n$ is the NLMS convergence factor defined in the interval $(0,1]$ and the parameter $\gamma$ is a constant introduced to avoid division by zero~\cite{benesty2010regularization}.

\subsection{RLS Algorithm}

The RLS algorithm aims to minimize, $\xi^d(k)$, the sum of the squared error (difference between the desired signal $d(k)$ and the filter output $y(k)$) using a recursive a method, given by
\begin{align}
    \xi^d(k) = \sum^k_{i=0} \lambda^{k-i} \left[ d(i) - {\bf x}^T(i) {\bf w}(k) \right]^2, 
\end{align}
where $\lambda$ is defined as an exponential weighting factor, ranging from $0 \ll \lambda \leq 1$, also called forgetting factor since the distant past effect is turned negligible with the constant updating (as $k$ increases). Despite the fast convergence, the RLS algorithm has higher computational complexity~\cite{diniz1997adaptive} and some stability problems in limited-precision implementations~\cite{cioffi1987limited,ardalan1986floating}. The RLS algorithm is
\begin{align}
    y(k) &= {\mathbf x}^T(k) {\mathbf w}(k-1), \\
    e(k) &= d(k) - y(k), \\
    {\mathbf \Psi}(k) &= {\mathbf S}_D(k-1) {\mathbf x}(k), \\
    {\mathbf S}_D(k) &= \frac{1}{\lambda} \left[ {\mathbf S}_D(k-1) - \frac{{\mathbf \Psi}(k) {\mathbf \Psi}^T(k)}{\lambda + {\mathbf \Psi}^T(k) {\mathbf x}(k)} \right], \label{eq:inv_Hessian} \\
    {\mathbf w}(k) &= {\mathbf w}(k-1) + e(k){\mathbf S}_D(k) {\mathbf x}(k), \label{eq:rls_update}
\end{align}
where ${\mathbf S}_D$ estimates the inverse Hessian, it is initialized with ${\mathbf S}_D(-1) = \delta{\mathbf I}$ and $\mathbf \Psi$ is an auxiliary column-vector introduced to reduce the computational burden~\cite{diniz1997adaptive}.

\section{The SOPOT Representation via MPGBP} \label{sec:SOPOTApprox}

The SOPOT numerical representation is similar to the well-known \textit{canonical-signed-digit} (CSD) representation.
The CSD representation consists of a ternary system that minimizes the number of non-zero digits by assuring that no non-zero CSD is adjacent to one another. In this representation, there is a higher probability of employing fewer non-zero digits (active bits) than in a usual 2's complement representation for $L$-bit word length~\cite{hewlitt2000canonical}.
However, in the SOPOT approximation, only a certain number of non-zero terms are kept, thus further reducing the number of active bits. The number of active bits defines the computational complexity of the multiplication operation. Since it bounds the number of shift operations required for the multiplication process, the greater the number of active bits, the greater the multiplication's computational burden.

\begin{figure}[hbt!]
    \centering
    \includegraphics[width=.9\linewidth]{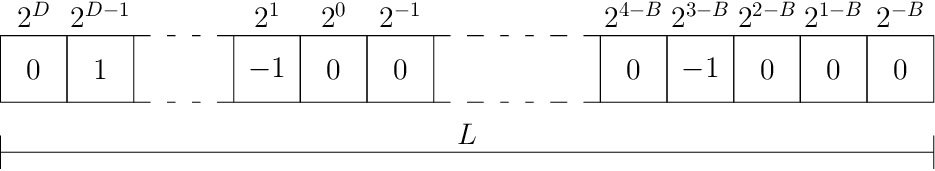}
    \caption{A SOPOT representation, where $L$ is the word-length, $2^D$ is the highest absolute value that an SPT term can assume, and $2^{-B}$ the lowest (in Equation (\ref{eq:sopot}), $\beta_l \in \{-B,~1-B,~\dots,~D-1,~D\}$.}
    \label{fig:sopot}
\end{figure}

The SOPOT approximation of a number $v$ is
\begin{align}
    v \approx \hat{a} = \sum^{L-1}_{l=0} \alpha_l 2^{\beta_l}, \label{eq:sopot}
\end{align}
where, $L$ is the word-length (bit-plane depth) of the numerical representation, $\alpha_l$ is $-1$, $0$ or $1$, and $\beta_l$ defines the bit-plane. Figure~\ref{fig:sopot} illustrates Equation (\ref{eq:sopot}). If ${\mathbf v}$ is an $N$-length vector ${\mathbf v} = [v_0~v_1~\cdots~v_{N-1}]^T$, the $n$-th element is approximated by the SOPOT,
\begin{align}
    \hat{v}_n = \sum^{L-1}_{l=0} \alpha_{l,n} 2^{\beta_{l,n}}. \label{eq:approx_vect_coef}
\end{align}

There are many techniques to obtain SOPOT approximations. But the MPGBP algorithm~\cite{da2014fir} presents the advantages of being fast and of low complexity to approximate vectors (a set of coefficients). At the same time, it allows imposing both the word length and, if desired, also a maximum number of non-null terms for the whole vector. In the MPGBP algorithm, described in Algorithm~\ref{alg:MPGBP}, a vector $\mathbf v$ is approximated as
\begin{align}
    {\mathbf v} \approx \sum_{i=1}^K 2^{k_i} {\mathbf c}_{j_i}, \label{eq:mpgbp_approx}
\end{align}
where ${\mathbf c}_j$ is a code-word from a dictionary $\mathcal{D} =\{ {\mathbf c}_1, ~{\mathbf c}_2, ~\dots, ~{\mathbf c}_j, ~\dots, ~{\mathbf c}_{\# \mathcal{D}} \}$ and $\# \mathcal{D}$ is the cardinality of the dictionary. In Equation (\ref{eq:mpgbp_approx}), $j_i$ is the selected code-word index in the $i$-th decomposition iteration. Defining code-words having coordinates -1, 0 and 1, one has that $2^{k_i} {\mathbf c}_{j_i}$ has coordinates that value $-2^{k_i}$, 0 or $2^{k_i}$. Furthermore, representations employing less terms are obtained if the number of non-null coordinates $P$ per code-word (${\mathbf c}_j$) equals $\lfloor \sqrt{N} \rfloor$~\cite{da2014fir}, the operator $\lfloor \cdot \rfloor$ returns the highest integer smaller than the operand. Thus, the MPGBP algorithm is a fast technique to approximate vectors (and matrices~\cite{mendoncca2019antenna}) successively. In our results, we use the MPGBP to obtain the initial representations of the input and filters for the adaptive processes.

Equation (\ref{eq:mpgbp_approx}) is a particular case of Equation (\ref{eq:approx_vect_coef}), the code-words determine the weights for all the vector coordinates in each bit-plane $2^{k_i}$ conjointly. Algorithm \ref{alg:MPGBP} illustrates this approximation process for the vector ${\mathbf v}$. One can set up the algorithm to use at most $M$ SPT terms (active bits limitation) or the deepest bit-plane value $B_\textrm{max}$ (the largest numeric precision).

Given that $2^k + 2^k = 2^{k+1}$, numeric representation in SOPOT are not unique. Since the objective is acquiring the approximation using the least SPT terms, a post-processing operation reduces the number of SPT terms. More details on this so-called \textit{common-terms reduction step} are presented in~\cite{da2014fir}. 

\begin{algorithm}
\caption{~~MPGBP for Vector Approximation}\label{alg:MPGBP}
\begin{algorithmic}[1]
\Statex {\textbf{input:} vector ${\mathbf v}$ of length $N$, and a maximum number of SPT terms $M_\textrm{max}$, maximum bit-depth $B_\textrm{max}$}
\Statex {\textbf{output:} the pairs of ${\mathbf c}_{j_i}$ (code-word) and $k_i$ (its associated bit-plane), the bit-depth $B$ and the number of total SPT terms $M$}
\Statex {\textbf{initialization:}} 
\State {$i\gets1$, ${\mathbf r}_1\gets{\mathbf v}$, $M\gets 0$, $B\gets 0$, $P=\left\lfloor \sqrt{N} \right\rfloor$}
\While{$M<M_\textrm{max}$ and $B<B_\textrm{max}$}
     \Statex {\textbf{code-word search:}}
     \State {Sort coordinates of ${\mathbf r}_i$ in decreasing order of their absolute values in $\mathcal{I}_P$}      
     \Statex  {\textbf{code-word update:}}
     \State {${\mathbf c}_{j_i} \leftarrow {\mathbf 0}$}
     \For{$p \in  \mathcal{I}_P$}     
     \If{$r_{j,p}>0$}
        \State $c_{j_i,p} \gets 1$ 
     \Else
         \State $c_{j_i,p} \gets -1$
     \EndIf
     \EndFor
     \Statex {\textbf{MPGBP data update:}}
     \State {$k_i \leftarrow \left \lceil - \log_2{\left ( \frac{4}{3P} \sum_{k \in \mathcal{I}_{P}}  |{r}_i(k)| \right )} \right \rceil$}
     \State {${\mathbf r}_{i+1} \gets {\mathbf r}_{i} - 2^{-k_i} {\mathbf c}_{j_i}$}
     \State {$i \gets i+1$} 
     \State {$B \gets \max_i{ {k_i}}$}
     \State {$M \gets \sum_{i}{\sum_n{|c_{j_i,n}|}}$}
\EndWhile
\State {\bf Common-terms reduction} (for details see~\cite{da2014fir})
\end{algorithmic}
\end{algorithm}

\section{SPT-Based Adaptive Filtering} \label{sec:spt_based}

Adaptive filters that use only the SOPOT arithmetic are evaluated. They are referred to as SPT-LMS (\textit{Signed Power-of-Two Least Mean Squares}), SPT-NLMS (\textit{Signed Power-of-Two Normalized Least Mean Squares}), and SPT-RLS (\textit{Signed Power-of-Two Recursive Least Squares}). The SOPOT representation imposes some numeric restrictions due to its finite accuracy and precision characteristics (accuracy is limited by the maximum number of active SPT terms and precision by the deepest bit plane). These characteristics grant the computational restriction. Thus the trade-off analysis is necessary. In our proposition, Algorithm~\ref{alg:MPGBP} is used to approximate the input signal in the tapped delay line, and the SOPOT arithmetic is preserved through the hole adaptive filtering process, including the coefficient update step. This implementation is better detailed in the following subsection.

\subsection{Implementation Aspects}

\subsubsection{Input Signal Representation}

If the input signal of an adaptive filter is represented correctly, then it is possible to obtain a reduction on the computational burden of the process~\cite{ameer2005multiplier,chen1996simplified,li1994multiplierless}. Furthermore, the SOPOT approximation via MPGBP is proposed in this work. Thus Algorithm~\ref{alg:MPGBP} applies to the tapped delay line that inputs the adaptive filters. Then, SOPOT arithmetic applies to the whole process.

\subsubsection{Multiplications}
Using the notation in equation \eqref{eq:sopot} one clearly sees that the multiplication of two SOPOT scalars $v'$ and $u'$ provides
\begin{align}
    v' u' = ~ & v_{0}u_{0}2^{b_{v_{0}}+b_{u_{0}}} +
    v_{0}u_{1}2^{b_{v_{0}}+b_{u_{1}}} + \cdots + \label{eq:multiply} \\
    +~&  v_{1}u_{0}2^{b_{v_{1}}+b_{u_{0}}} + \cdots + 
    v_{M-1}u_{M-1}2^{b_{v_{M-1}}+b_{u_{M-1}}} \notag
\end{align}
and the multiplication process turns into additions of SPT terms exponents. In order for the multiplication of two SOPOT-represented scalars having the same amount of SPT terms as its predecessors, the $M$ highest exponents resulting from equation (\ref{eq:multiply}) are retained.

\subsubsection{Inner Product} An inner product is the sum of element-wise multiplications of vectors. Therefore, this element-wise multiplication is defined as the key-operation in the inner and matrix products. As one observes from (\ref{eq:lms_output}) to (\ref{eq:rls_update}), the LMS, NLMS and RLS algorithms require computing inner products between matrices and vectors. Let ${\mathbf v}' = [v(0)', v(1)', \dots, v(N-1)']^T$ and ${\mathbf u}'=[u(0)', u(1)', \dots, u(N-1)']^T$ be two SOPOT-represented line vectors of length $N$, then ${\mathbf v}'^T{\mathbf u}'$ accumulates $N$ sums as the one in equation (\ref{eq:multiply}). The product of $N\times N$ matrix by an $N$-length vector and the product between two $N\times N$ matrices are implemented using $N$ and $N^2$ inner products, respectively.

\subsubsection{Denominator Inversion}
Equations \eqref{eq:nlms_update} and \eqref{eq:inv_Hessian} require division operations for the NLMS and RLS algorithms. Performing division when the denominator is in SOPOT is not trivial. Therefore, we propose to approximate the inverted denominator and obtain the division by means of a multiplication. Using a limited number of the $l$-terms in the Maclaurin Series for $v$
\begin{align}
    v^{-1} = \sum^\infty_{l=0} (1-v)^l,~ \{ v \in \mathbb{R}| -2<v<2 \},
\end{align}
one can approximate the inversion for a given denominator $v$. If the denominator is not in the valid interval of the Maclaurin Series, we propose to consider only the most significant SPT term for the inversion, i.e., for a given scalar $v=\pm 2^i \pm 2^j \pm \cdots \pm 2^k$, with $i>j>k$, the approximation results in $v^{-1} \approx \pm 2^{-i}$, the greater the difference between $i$ and $j$, the better this approximation is.

\subsubsection{Complexity Constraint}

One can limit the word length $L$ by limiting $B$ (which denotes the lowest exponent of the sum) and the quantity $M$ of SPT terms. These restrictions are key elements for the complexity reduction and the representation precision, therefore having a major influence on the quantization error. For example, when representing a value, $B$ limits the lowest exponent of the SPT term to be an integer within $[D,-B]$, $D$ limits its highest exponent, and $L$ is defined by the equation $L = D + B - 1$. Although, in each operation, the least significant SPT terms are ignored, keeping only the $M$ largest terms. This truncation maintains the complexity restriction but may produce numerical errors on the adaptive filtering process~\cite{cioffi1987limited,ardalan1986floating}. Therefore, to measure the complexity of different SOPOT approximations of an $N$-length filter, we define $M/N$, the ratio between the number of active SPTs and the filter length.

\subsection{SOPOT Quantization Effects on Adaptive Filters}\label{sec:quant_effect}

Aiming to certify that the SOPOT quantization used in the RLS algorithm does not jeopardize stability, some analyses are presented that evaluate the impact of the numerical restriction of the approximations in the adaptive filter.

\subsubsection{Limited-precision Effects}

Quantized adaptive filtering implementations are only possible if the quantization error has zero mean~\cite{ardalan1986floating,cioffi1987limited}. For some quantization methods, this result is trivial, although it is not evident for the SOPOT implementation, and thus, some analyses are required.

For a given vector source ${\mathbf v} \in \mathcal{V}$, the expected value of the quantization error ${\mathbf e}_q$ is
\begin{align}
    E[{\mathbf e}_q] = E[{\mathbf v} - \hat{\mathbf v}].
\end{align}
If the expected value for each coordinate in $\mathbf v$ is zero and the vector $\mathbf v$ is approximated using the MPGBP algorithm, then
\begin{align}
    E[{\mathbf e}_q] = -E \left[ \sum_i 2^{\beta_i} {\mathbf c}_i \right],
\end{align}
where each ${\mathbf c}_i$ has $P = \lfloor \sqrt{N} \rfloor$ non zero coordinates (consisting of $-1$ and $1$). If the probability of the codewords coefficients is the same independently of the exponent $\beta_i$, i.e., independently of the magnitude $2^{\beta_i}$ of the vector to be approximated~\cite{lovisolo2010statistics}, thus
\begin{align}
    E[{\mathbf e}_q] = - \sum_i E\left[2^{\beta_i} \right] E[{\mathbf c}_i]. \label{eq:error_bound}
\end{align}
By definition/construction of the codewords, $E[{\mathbf c}_i] = {\mathbf 0}$ if the vector source $\mathcal{V}$ is uniformly distributed in $\mathbb{R}^N$~\cite{lovisolo2010statistics}.

A numerical analysis is now presented. In this analysis, the MPGBP algorithm is used to approximate a series of $N$-length random vectors following Gaussian (zero mean and unitary variance) distribution using at most $M$ SOPOT in the vector approximation. The process is repeated exhaustively ($ensemble = 10^5$) in a Monte Carlo method. Table \ref{tab:expect_val} presents some statistical analyses of the quantization error for different lengths and complexity constraints. Endorsed by the Law of Large Numbers (LLN), one can estimate the mean vector of the distribution using,
\begin{align}
    E[{\mathbf e}_q] = \mu_{{\mathbf e}_q} = \frac{1}{10^5} \sum^{10^5}_{k=1} ({\mathbf v} - \hat{\mathbf v})_k,
\end{align}
where $k$ indexes the trials. In Table~\ref{tab:expect_val} the expected values correspond to the largest absolute value among the means of the coordinates of ${\mathbf e}_q$. The variance vector is estimated using
\begin{align}
    {\mathbf \sigma}^2_{{\mathbf e}_q} = E\left[ \left( {\mathbf e_q} - E[{\mathbf e}_q] \right)^2 \right].
\end{align}
Table~\ref{tab:expect_val} also presents the more significant value among the coordinates of the variance vector.

\begin{figure}[hbt!]
    \centering
    \includegraphics[width=\linewidth]{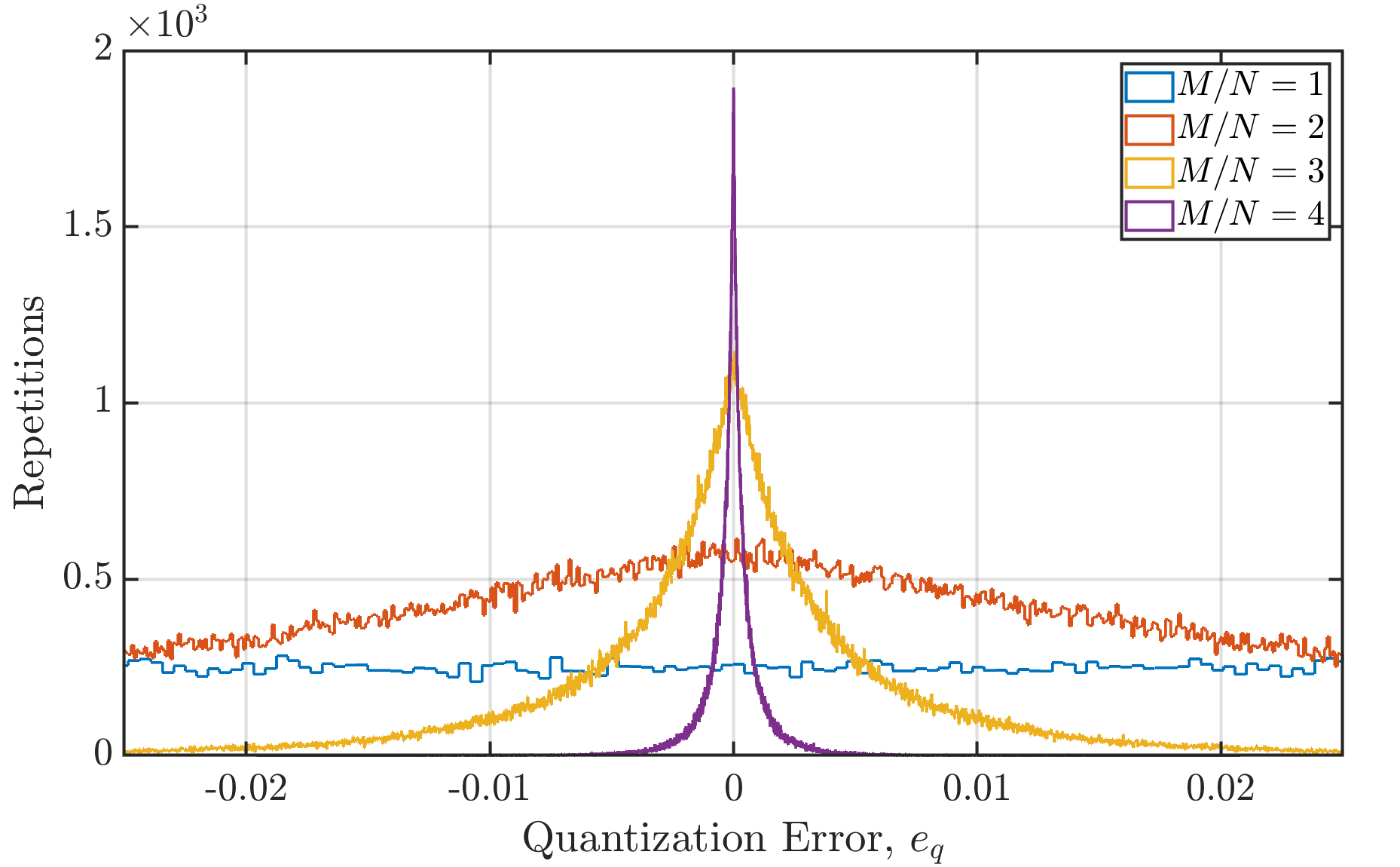}
    \caption{Histogram, with 10000 bins, of the quantization error originated from the SOPOT approximation using the MPGBP algorithm. Considering the approximation of a 4-element vector source with $M/N=1,~2,~3 \textrm{ and } 4$.}
    \label{fig:hist}
\end{figure}

The values in Table~\ref{tab:expect_val} are very small, and the more SPT terms, the smaller the variance. These results suggest that the MPGBP algorithm is capable of approximating vectors without bias. Figure~\ref{fig:hist} presents the distribution of the quantization error in each element of a 4-element ($N=4$) vector source stemming from SOPOT approximation using the MPGBP algorithm with complexities $M/N=1,~2,~3 \textrm{ and } 4$. For high restraint in complexity ($M/N<3$), one can observe that the distribution tends to a uniform, and the mean does not tend to zero. However, as we allow more SPT terms, the quantization error mean approaches zero, and the variance decreases.

\begin{table}[t]
\caption{Expected value ($\mu$) and variance ($\sigma^2$) of the quantization error stemming from the MPGBP approximation algorithm.}
\label{tab:expect_val}
\centering
\begin{footnotesize}
\begin{tabular}{|c|c|c|c|c|}
\cline{3-5}
\multicolumn{2}{c}{} & \multicolumn{3}{|c|}{$N$} \\ \hline
$(M/N)$ & & $4$ & $12$  & $16$  \\ \hline
\multirow{2}{*}{3} & $\mu_{{\bf e}_q}$ &  $1.99 \times 10^{-5}$ & $1.29 \times 10^{-5}$ & $2.06 \times 10^{-5}$ \\ \cline{2-5}
& $\sigma^2_{{\bf e}_q}$ & $ 7.02 \times 10^{-5}$ & $1.22 \times 10^{-5}$ & $1.03 \times 10^{-5}$ \\ \hline 
\multirow{2}{*}{5} & $\mu_{{\bf e}_q}$ &  $1.02 \times 10^{-6}$ & $3.33 \times 10^{-7}$ & $3.74 \times 10^{-7}$ \\ \cline{2-5}
& $\sigma^2_{{\bf e}_q}$ & $ 3.48 \times 10^{-8}$ & $4.09 \times 10^{-9}$ & $3.27 \times 10^{-9}$ \\ \hline 
\multirow{2}{*}{7} & $\mu_{{\bf e}_q}$ &  $1.58 \times 10^{-8}$ & $6.92 \times 10^{-9}$ & $9.10 \times 10^{-9}$ \\ \cline{2-5}
& $\sigma^2_{{\bf e}_q}$ & $ 1.74 \times 10^{-11}$ & $1.30 \times 10^{-12}$ & $9.78 \times 10^{-13}$ \\ \hline 
\multirow{2}{*}{10} & $\mu_{{\bf e}_q}$ &  $8.78 \times 10^{-11}$ & $2.29 \times 10^{-11}$ & $1.38 \times 10^{-11}$ \\ \cline{2-5}
& $\sigma^2_{{\bf e}_q}$ & $ 1.89 \times 10^{-16}$ & $7.78 \times 10^{-18}$ & $5.24 \times 10^{-18}$ \\ \hline 
\end{tabular}
\end{footnotesize}
\end{table}

\subsubsection{Maximum Complexity Restriction}

In order for the adaptive filter to converge, the zero-mean quantization error must be kept throughout the iterations. To constraint complexity,  approximations are truncated after every operation, preserving only the $M$ most significant SPT terms of the SOPOT. Arguably, the elimination of SPT terms that are randomly negative or positive does not change the zero-mean quantization error. Since i) the construction of the codewords ${\mathbf c}_i$ sustains both values $-1$ or $1$ in every coordinate; and ii) for a uniformly distributed vector source, the residues of each MPGBP iteration result in symmetric distributions. Therefore, it is expected that the complexity constraint spreads the quantization error distribution shown in Figure~\ref{fig:hist} without changing its mean.

\section{Simulation Results}\label{sec:results}

The proposed SOPOT adaptive filters are applied in two different scenarios. Firstly, the SOPOT filters are implemented, along with fixed-point (another hardware known for its low computational complexity) filters, in a classical system identification scenario (Figure~\ref{fig:SysID}), it is intended to evaluate gaps in the performance between both implementations. Secondly, a change detection application evaluates the impact of the approximation on the detector's performance.

\subsection{System Identification Application}

\begin{figure}[hbt!]
    \centering
    \includegraphics[width=\linewidth]{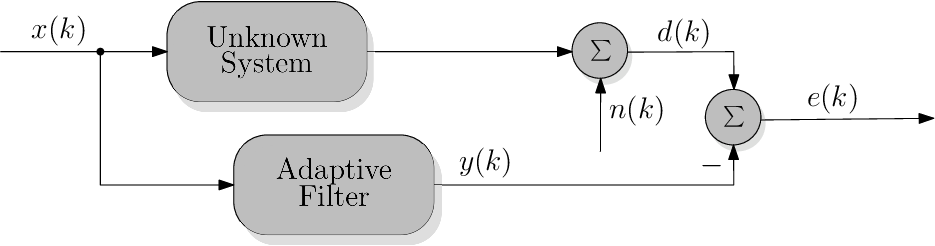}
    \caption{System identification scenario block diagram: the adaptive filter tries to mimic the effects of the unknown system in $x(k)$ and therefore can be used to identify it.}
    \label{fig:SysID}
\end{figure}

In this application, adaptive filters are used to identify a randomly generated normalized vector, which represents a FIR filter and is used as the \textit{unknown system}. This classical application can help evaluating the filter's convergence according to the imposed complexity restraint and the achievable accuracy in the system identification problem. As in Figure~\ref{fig:SysID}, the input signal $x(k)$ is a vector following a Gaussian distribution. The additive noise $n(k)$ is Gaussian white noise, with adjustable SNR. The adaptive filter will try to mimic the unknown system's effects in the input signal, and therefore the output signal $y(k)$ is an estimate for $d(k)$. Since $x(k)$ and $n(k)$ are uncorrelated, the adaptive filter is not able to nullify its effects, thus the error signal $e(k)$ is limited by the SNR in its reduction. The filters are of $3$rd order, i.e., they are also $4$-elements vectors, $N=4$. The SNR is set by adjusting the noise power as follows
\begin{align}
    n(k) = \sqrt{\frac{\left(\frac{1}{K}\sum_{k=1}^K |x(k)|^2 \right) 10^{-\textrm{SNR}_{\textrm{dB}}/10}}{\frac{1}{K} \sum_{k=1}^K |n'(k)|^2}} ~ n'(k),
\end{align}
where $\textrm{SNR}_{\textrm{dB}}$ is the desired signal to noise ratio, in dB,  $n'(k)$ is the misadjusted noise signal, and the noise signal $n(k)$ is the adjusted noise signal.

The adaptive filters are exhaustively trained with $10000$ (ten thousand) iterations each, and a Monte Carlo method is applied with $150$ repetitions. The experiments require configuring the parameters of the adaptive filters. The parameters for the SPT-based implementation are shown Table~\ref{tab:parameter_set}, while for the fixed-point the parameters are $\mu = 0.075$, $\mu_n = 0.13$, $\gamma=1e^{-4}$, $\lambda=0.9987$ and $\delta=0.06$. The SPT filters and the fixed-point filters use different values for $\lambda$ in the RLS algorithm. The difference arises from the need to have all values expressed as SOPOTs for the SPT filters. For both the SPT and fixed-point algorithms, the parameter values were empirically adjusted for best performance, i.e., the smallest MSE (equation~\eqref{eq:mse:_def}). We note that this parameters are for fixed step-size algorithms; further work is necessary to tune them for variable step-size algorithms.

\begin{table*}[ht!]
    \centering
    \caption{Parameters used in the SPT-based adaptive filter algorithms. The original floating-point values are shown in parentheses.}
    \begin{footnotesize}
    \begin{tabular}{|c|c|c|c|c|c|} \hline
        \multirow{2}{*}{$M/N$}& \multicolumn{1}{|c|}{LMS} &  \multicolumn{2}{|c|}{NLMS} &   \multicolumn{2}{|c|}{RLS} \\ \cline{2-6}
        & $\mu$& $\mu_n$ & $\gamma$ & $\lambda$ & $\delta$ \\ \hline
        $1$& $2^{-4} ~ (0.075)$ & $2^{-3} ~ (0.13)$ & $0 ~ (1e^{-4})$& $2^0 ~ (0.975)$ & $2^{-1} ~ (0.06)$ \\ \hline
        $2$& $2^{-4} + 2^{-6} ~ (0.075)$ & $2^{-3} + 2^{-8}  ~ (0.13)$ & $0 ~ (1e^{-4})$ & $2^0 - 2^{-5} ~ (0.975)$ & $2^{-1} + 2^{-3} ~ (0.06)$ \\ \hline
        $3$& $2^{-4} + 2^{-6} - 2^{-8} ~ (0.075)$ & $2^{-3} + 2^{-8} + 2^{-10} ~ (0.13)$ & $0 ~ (1e^{-4})$ & $2^0 - 2^{-5} - 2^{-7} ~ (0.96)$ & $2^{-1} + 2^{-3} - 2^{-5} ~ (0.06)$ \\ \hline
        $4$& \begin{tabular}[c]{@{}c@{}} $2^{-4} + 2^{-6} - 2^{-8}$ \\$+ 2^{-10} ~ (0.075)$ \end{tabular} & $2^{-3} + 2^{-8} + 2^{-10} ~ (0.13)$ & $0 ~ (1e^{-4})$ & \begin{tabular}[c]{@{}c@{}}$2^0 - 2^{-5} - 2^{-7}$\\$- 2^{-10} ~ (0.96)$\end{tabular} & \begin{tabular}[c]{@{}c@{}}$2^{-1} + 2^{-3} - 2^{-5}$\\$ + 2^{-7} ~ (0.06)$\end{tabular} \\ \hline
        $5$& \begin{tabular}[c]{@{}c@{}} $2^{-4} + 2^{-6} - 2^{-8}$ \\$+ 2^{-10} - 2^{-12} ~ (0.075)$\end{tabular} & $2^{-3} + 2^{-8} + 2^{-10} ~ (0.13)$ & $0 ~ (1e^{-4})$ & \begin{tabular}[c]{@{}c@{}}$2^0 - 2^{-5} - 2^{-6}$\\$ + 2^{-9} ~ (0.955)$\end{tabular} & \begin{tabular}[c]{@{}c@{}}$2^{-1} + 2^{-3} - 2^{-5}$\\$ + 2^{-7} - 2^{-9} ~ (0.06)$\end{tabular} \\ \hline
        $6$& \begin{tabular}[c]{@{}c@{}} $2^{-4} + 2^{-6} - 2^{-8}$\\$ + 2^{-10} - 2^{-12} ~ (0.075)$\end{tabular} & $2^{-3} + 2^{-8} + 2^{-10} ~ (0.13)$ & $0 ~ (1e^{-4})$ & \begin{tabular}[c]{@{}c@{}}$2^0 - 2^{-5} - 2^{-6}$\\$ + 2^{-10} ~ (0.954)$\end{tabular} & \begin{tabular}[c]{@{}c@{}}$2^{-1} + 2^{-3} - 2^{-5}$\\$ + 2^{-7} - 2^{-9} + 2^{-11} ~ (0.06)$\end{tabular} \\ \hline
        $7$& \begin{tabular}[c]{@{}c@{}}$2^{-4} + 2^{-6} - 2^{-8}$\\$ + 2^{-10} - 2^{-12} ~ (0.075)$\end{tabular} & $2^{-3} + 2^{-8} + 2^{-10} ~ (0.13)$ & $0 ~ (1e^{-4})$ & \begin{tabular}[c]{@{}c@{}}$2^0 - 2^{-5} - 2^{-7}$\\$ - 2^{-10} ~ (0.96)$\end{tabular} & \begin{tabular}[c]{@{}c@{}}$2^{-1} + 2^{-3} - 2^{-5}$\\$ + 2^{-7} - 2^{-9} + 2^{-11} ~ (0.06)$\end{tabular} \\ \hline
        $8$& \begin{tabular}[c]{@{}c@{}}$2^{-4} + 2^{-6} - 2^{-8}$\\$ + 2^{-10} - 2^{-12} ~ (0.075)$\end{tabular} & $2^{-3} + 2^{-8} + 2^{-10} ~ (0.13)$ & $0 ~ (1e^{-4})$ & \begin{tabular}[c]{@{}c@{}}$2^0 - 2^{-5} - 2^{-6}$\\$ + 2^{-10} ~ (0.954)$\end{tabular} & \begin{tabular}[c]{@{}c@{}}$2^{-1} + 2^{-3} - 2^{-5}$\\$ + 2^{-7} - 2^{-9} + 2^{-11} ~ (0.06)$\end{tabular} \\ \hline
    \end{tabular}
    \end{footnotesize}
    \label{tab:parameter_set}
\end{table*}

\subsubsection{Performance Evaluation}

To evaluate the algorithm's convergence after the numeric approximation, one can evaluate the mean squared error (MSE) curve, which is an estimate for the MSE using the filter's error signal through the Monte Carlo process. This curve is calculated as follows,
\begin{align}
    \textrm{MSE}(k) = 10 \log_{10} \left(\frac{\sum^{150}_{i=1} e^2_i(k)}{150}\right), \label{eq:mse:_def}
\end{align}
where $e_i(k)$ is the error signal for $k$-th iteration of the $i$-th Monte Carlo repetition (there are 150 repetitions).

\begin{figure}[hbt!]
    \centering
    \includegraphics[width=\linewidth]{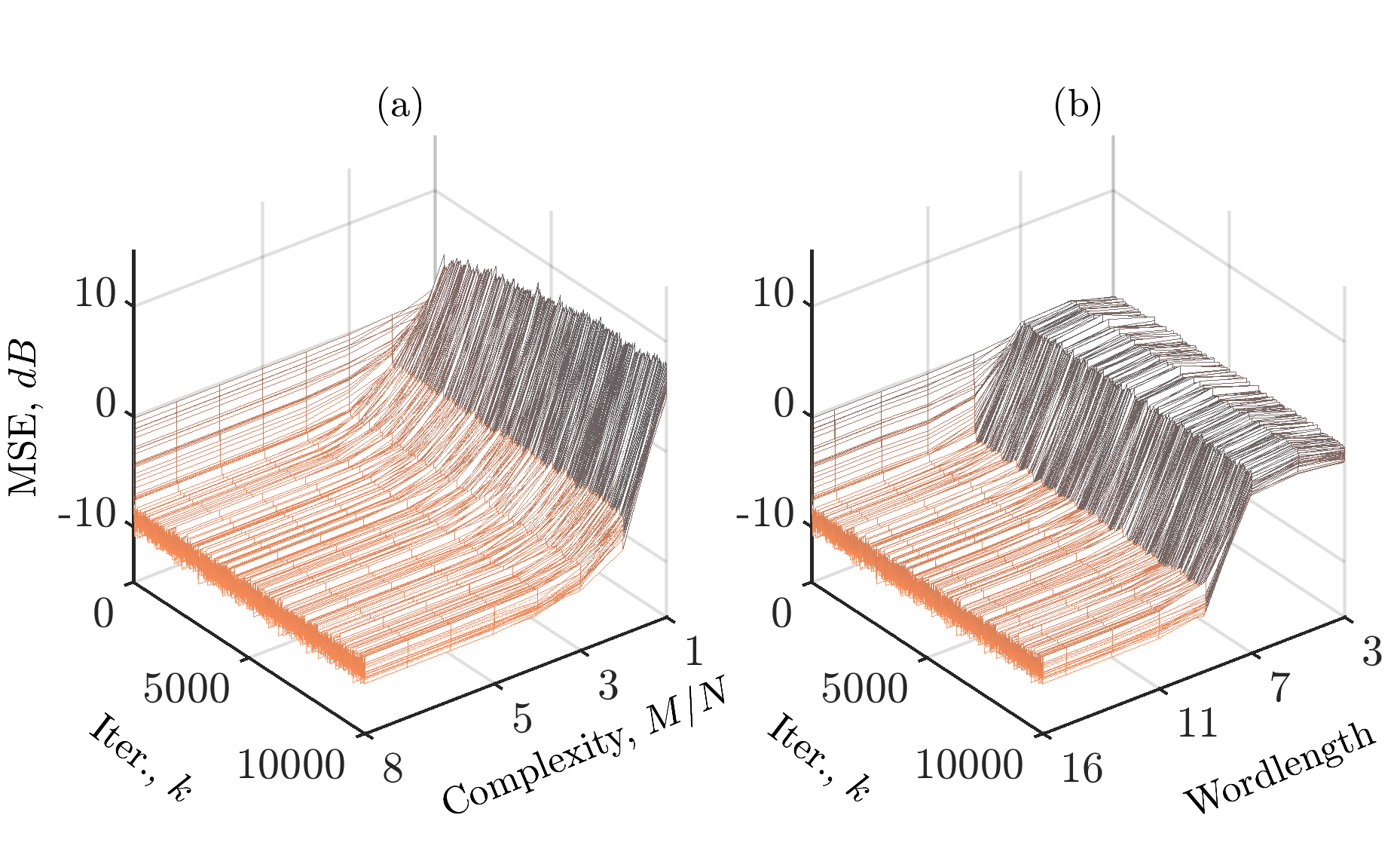}
    \caption{MSE curve evolution with $\textrm{SNR}=10~dB$ (a) for different complexities, $M/N$, in the SPT-LMS and (b) for different word lengths in the fixed-point LMS.}
    \label{fig:MSE_LMS}
\end{figure}

\begin{figure}[hbt!]
    \centering
    \includegraphics[width=\linewidth]{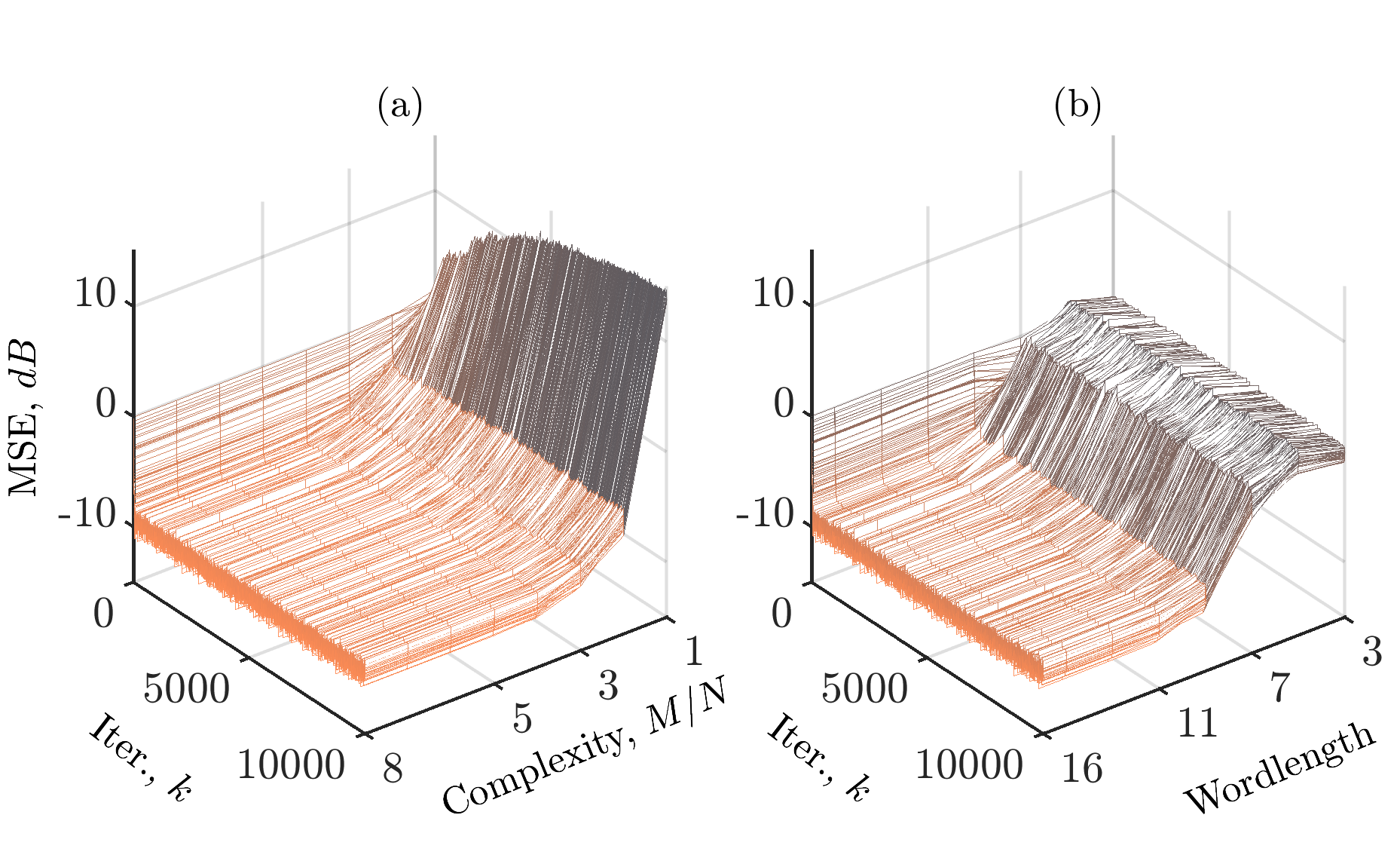}
    \caption{MSE curve evolution with $\textrm{SNR}=10~dB$ (a) for different complexities, $M/N$, in the SPT-NLMS and (b) for different word lengths in the fixed-point NLMS.}
    \label{fig:MSE_NLMS}
\end{figure}

\begin{figure}[hbt!]
    \centering
    \includegraphics[width=\linewidth]{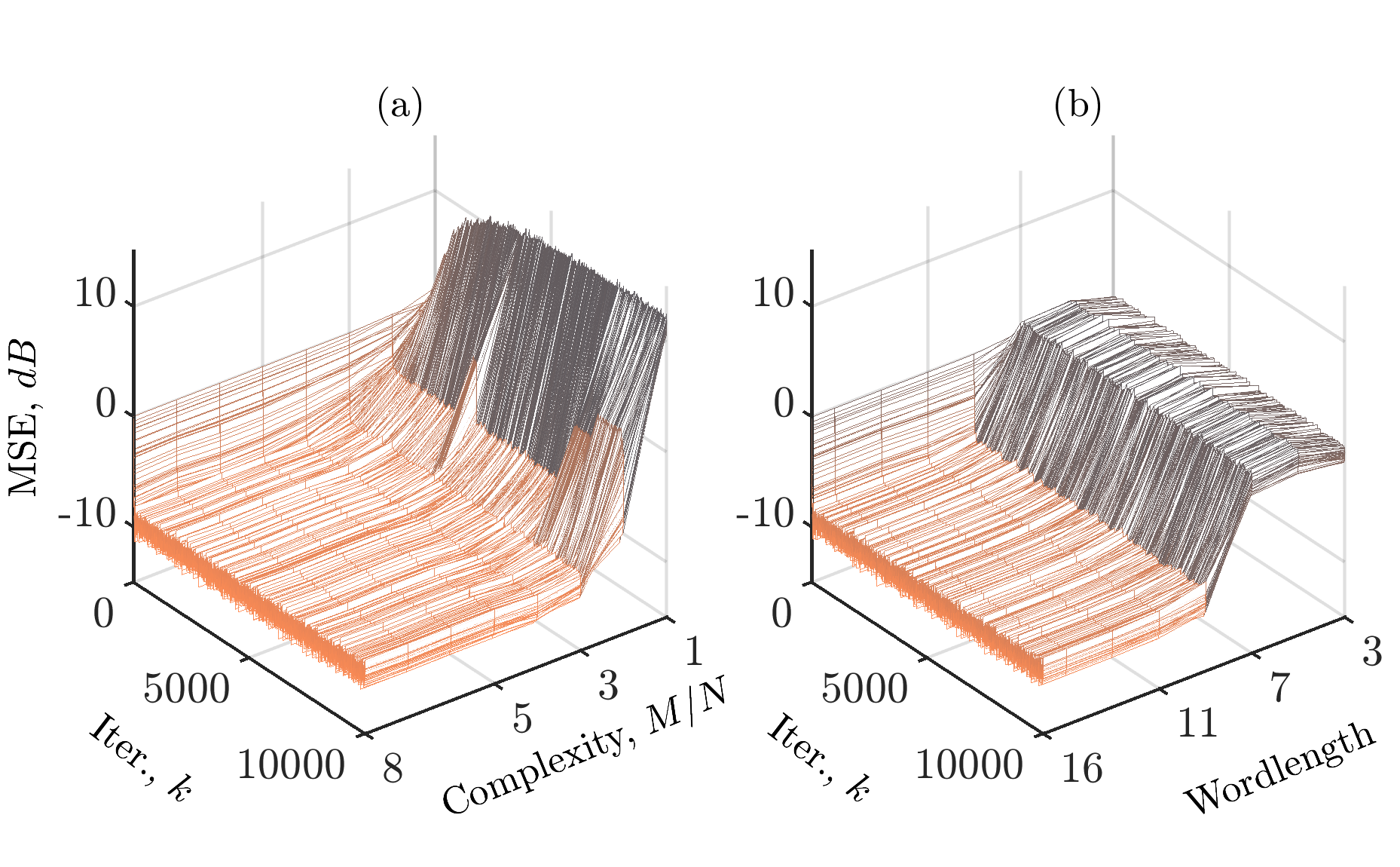}
    \caption{MSE curve evolution with $\textrm{SNR}=10~dB$ (a) for different complexities, $M/N$, in the SPT-RLS and (b) for different word lengths in the fixed-point RLS.}
    \label{fig:MSE_RLS}
\end{figure}

The MSE curves of SPT-LMS, SPT-NLMS and SPT-RLS algorithms for different SOPOT complexities ($M/N$) and fixed-point word lengths are presented in Figures~\ref{fig:MSE_LMS}, \ref{fig:MSE_NLMS} and \ref{fig:MSE_RLS}, respectively. In this case, filters are trained for $\textrm{SNR}=10~\textrm{dB}$. Observe that these curves are two-dimensional since, besides complexity or word length, there is also the iteration axis. They allow us to verify that the error signal evolution is stable throughout the iterations, i.e., the filter does not diverge through the entire process. One may also notice that the fixed-point performance degrades in word lengths under 7 bits, while the SPT variation can already perform quite well with only 3 SPT terms.

The MSD is calculated as the average squared distance between the filter coefficients $\ell_2$-norm and the unknown system's coefficients $\ell_2$-norm through the Mote Carlo process. It presents the deviation between adaptive filters and unknown system coefficient norm and can be calculated as
\begin{align}
    \textrm{MSD}(k) = 10 \log_{10} \left( \frac{\sum_{i=1}^{150} \left(||{\mathbf w}_i(k)||_2 - ||{\bf h}||_2 \right)^2 }{150} \right),
\end{align}
where ${\mathbf w}_i(k)$ is a $N$-element vector representing the adaptive filter coefficients in the $k$-th iteration for the $i$-th repetition of the Monte Carlo and $\bf h$ is the $4$-element vector representing the \textit{unknown system}'s impulse response.

\begin{figure}[hbt!]
    \centering
    \includegraphics[width=\linewidth]{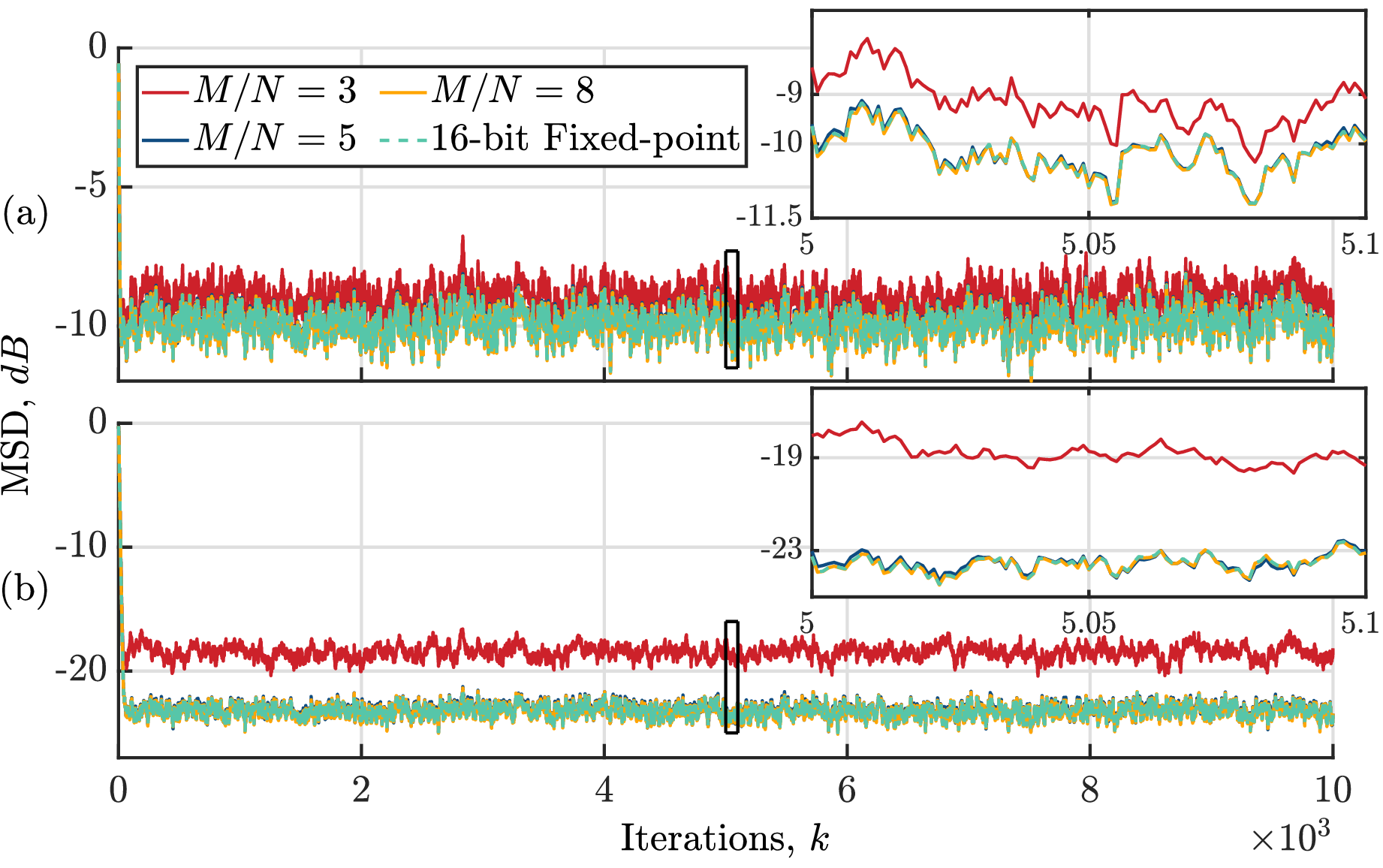}
    \caption{MSD for the SPT-LMS with different complexities, $M/N$ (SPT terms per coefficient), compared with a $16$-bit fixed-point LMS, for (a) $\textrm{SNR}=-3~dB$ and (b) $\textrm{SNR}=10~dB$.}
    \label{fig:LMS_precision}
\end{figure}

\begin{figure}[hbt!]
    \centering
    \includegraphics[width=\linewidth]{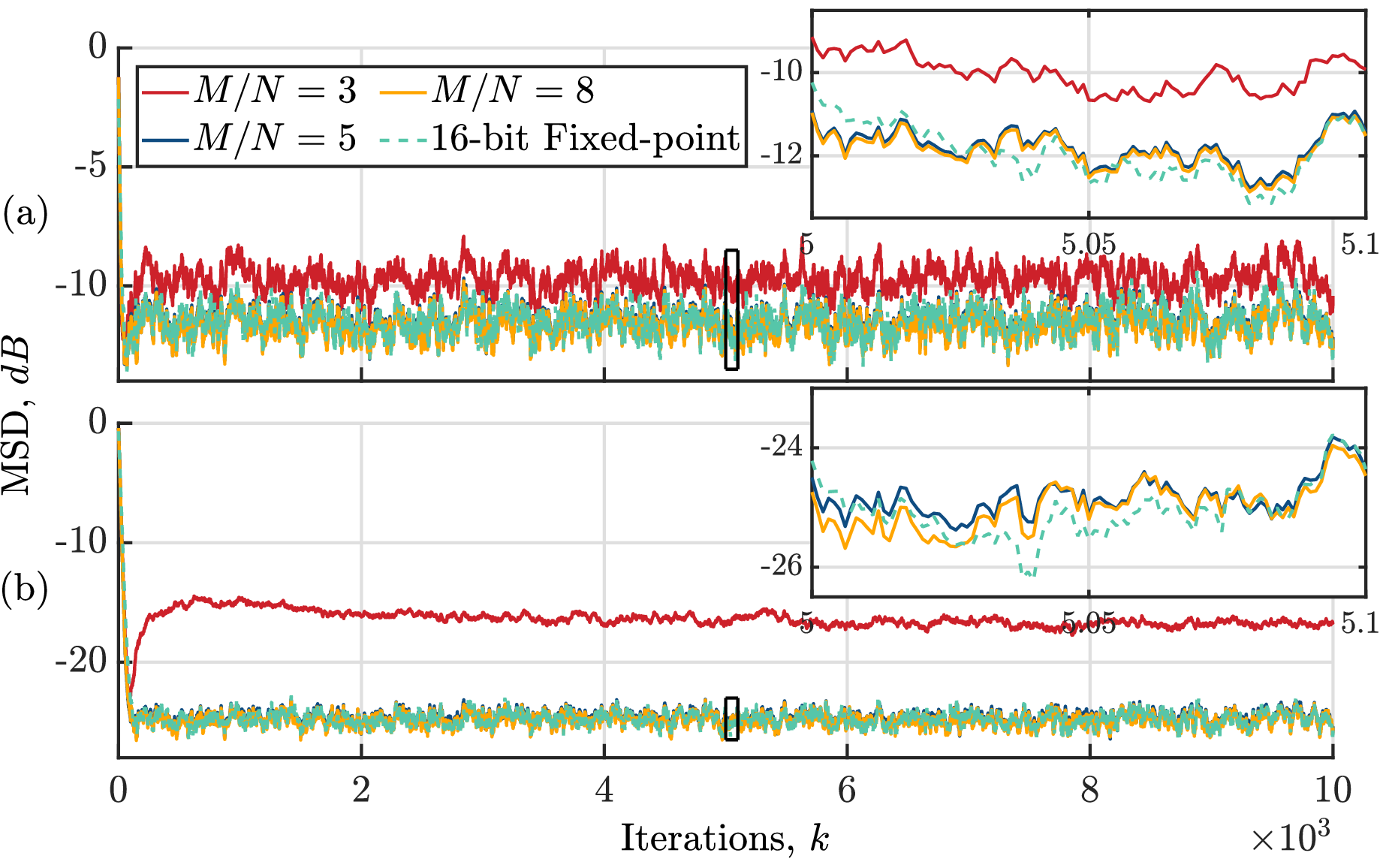}
    \caption{MSD for the SPT-NLMS with different complexities, $M/N$ (SPT terms per coefficient), compared with a $16$-bit fixed-point NLMS, for (a) $\textrm{SNR}=-3~dB$ and (b) $\textrm{SNR}=10~dB$.}
    \label{fig:NLMS_precision}
\end{figure}

\begin{figure}[hbt!]
    \centering
    \includegraphics[width=\linewidth]{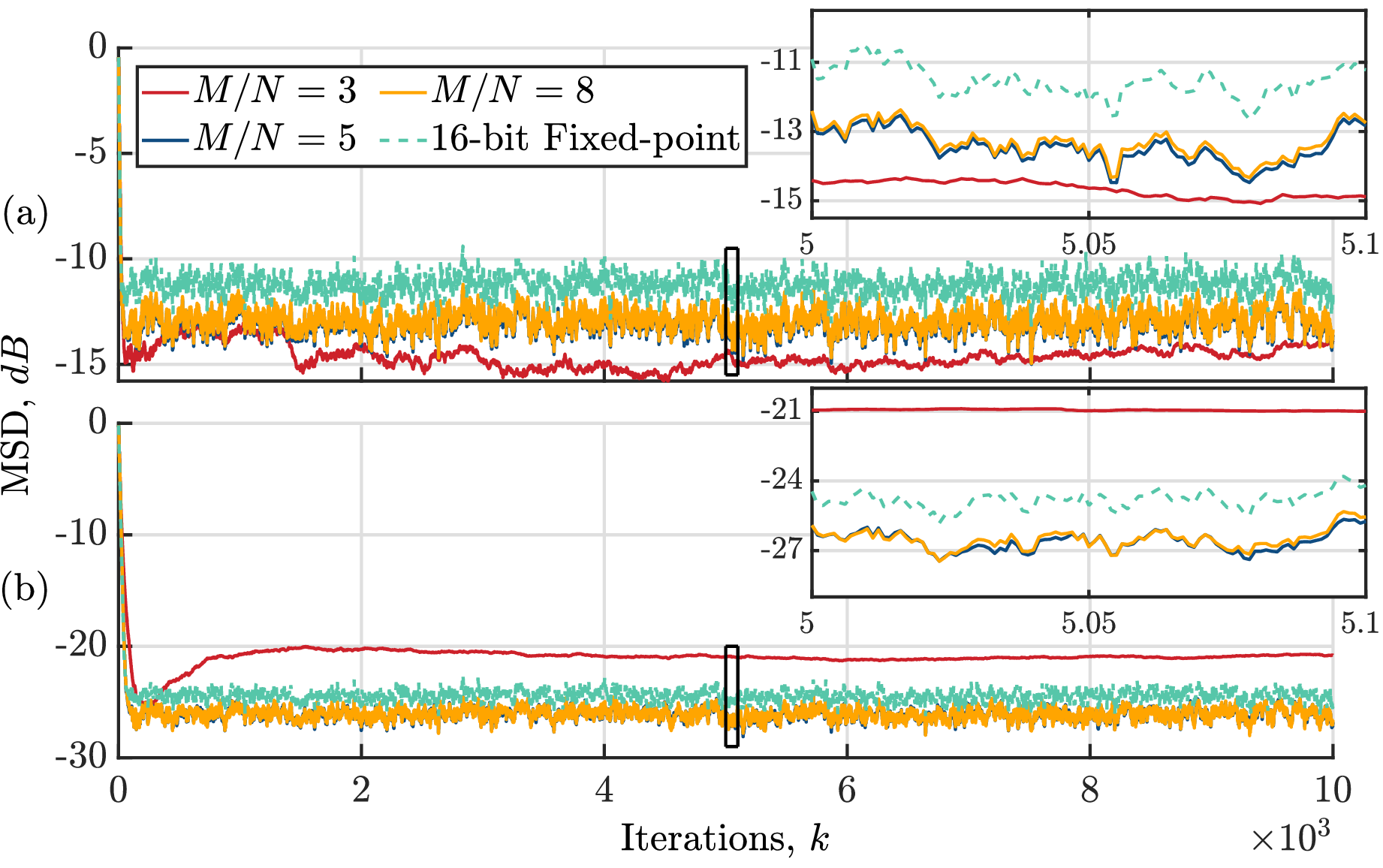}
    \caption{MSD for the SPT-RLS with different complexities, $M/N$ (SPT terms per coefficient), compared with a $16$-bit fixed-point RLS, for (a) $\textrm{SNR}=-3~dB$ and (b) $\textrm{SNR}=10~dB$.}
    \label{fig:RLS_precision}
\end{figure}

Figures~\ref{fig:LMS_precision},~\ref{fig:NLMS_precision} and \ref{fig:RLS_precision} compare the SPT-based adaptive filters with different complexities ($M/N$) with a fixed-point of 16-bit fixed-point precision. In this case, filters are trained for $\textrm{SNR}=-3 \textrm{ and } 10~\textrm{dB}$. These results compare the capability of the different SOPOT complexity setups to represent numbers. Despite the limitation in the number of SPT terms, the active terms can move freely through the allowed bit-plane range enabling higher numerical precision with low computational complexity implementations. Filters with lower complexity present slightly inferior performance, but this difference is mitigated under low SNR.

\subsubsection{SPT-based and Fixed-Point Filters on Equal Terms}

The complexity of the RLS algorithm is proportional to $N^2$. On the other hand, the complexities of the LMS and NLMS algorithms are proportional to $2N$ and $3N$, respectively.

Let the fixed-point word length be $B_F$ bits. The multiplication between two fixed-point numbers requires that one be shifted depending on the non-null terms of the other. Then, the resulting shifted binary words must be added, resulting in a $2B_F$-bit word (rounded accordingly to return a $B_F$-bit word). Assuming a sufficient long word and that the encoded values are distributed symmetrically over the fixed-point range, the average number of active bits per fixed-point word is $B_F/2$. This reduces the complexity of the multiplication between two fixed-point numbers to $B_F/2$ shifts of $B_F$-bit words and $B_F/2 - 1$ additions of possibly $2B_F$-bit words.

On the other hand, a SOPOT representation using $M$ SPT terms and $B$-bit exponents requires at most $M^2$ additions (distributive property) of $B$-bit words -- this is a worst-case scenario analysis -- for the same multiplication operation. Regardless of storage requirements, one may compare the complexities of both representations for the multiplication of two numbers using the amount of necessary ``$1$-bit sums''. Consequently, the fixed-point multiplication uses $B_F^2 - 2B_F$ sums while the SOPOT multiplication uses $BM^2$ sums.

Therefore, to evaluate SOPOT and fixed-point filters under similar computational complexities, the value 
\begin{equation}
B_F = 1+\sqrt{1+BM^2} \approx \left\lceil 1+M\sqrt{B} \right\rceil
\end{equation}
presents the fixed-point word length counterpart for a given number of SPT terms $M$.

\begin{figure}[hbt!]
    \centering
    \includegraphics[width=\linewidth]{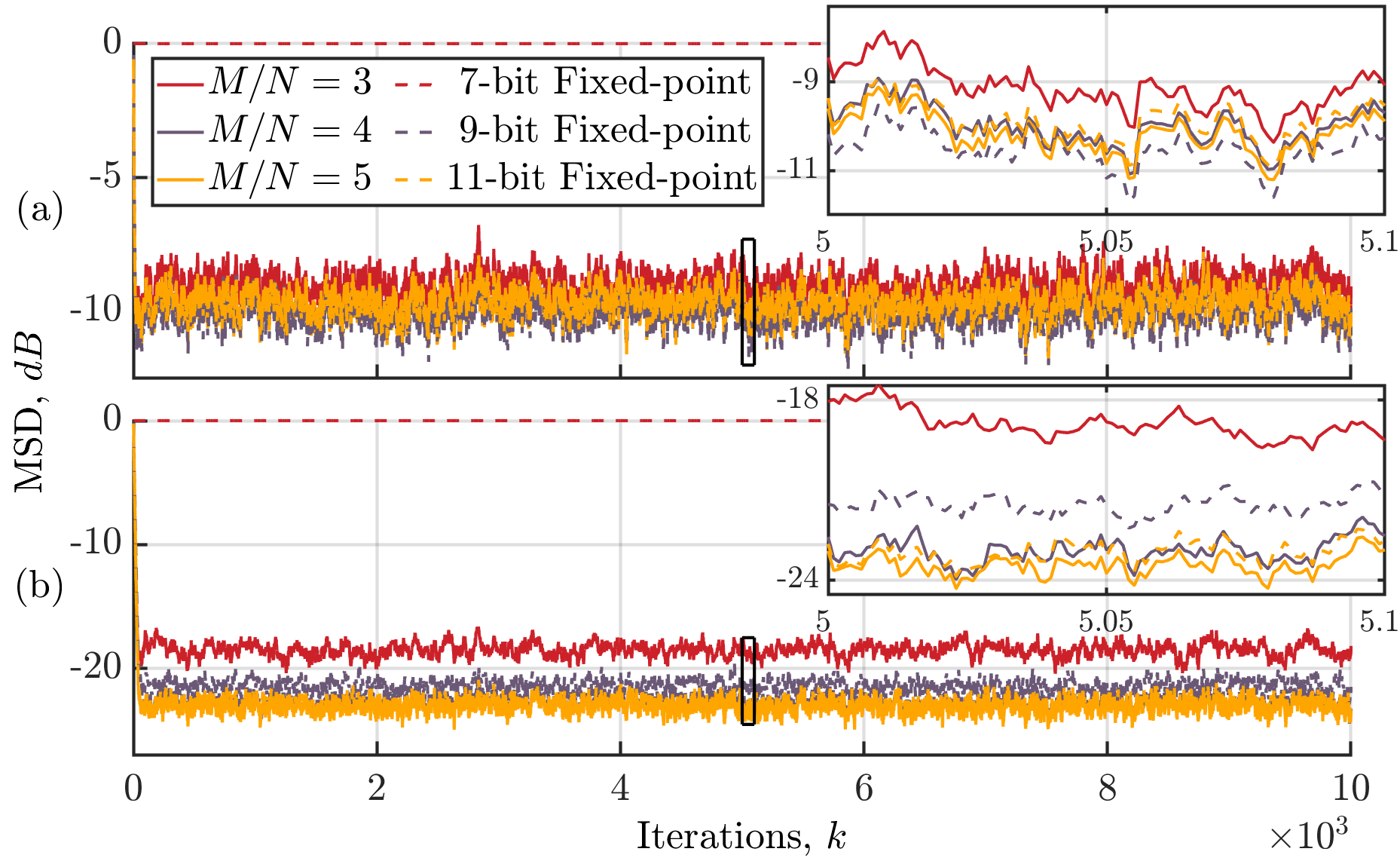}
    \caption{MSD for the SPT-LMS with different complexities, $M/N$ (SPT terms per coefficient) compared with fixed-point of estimated similar complexity, for (a) $\textrm{SNR}=-3~dB$ and (b) $\textrm{SNR}=10~dB$.}
    \label{fig:LMS_complexity}
\end{figure}

\begin{figure}[hbt!]
    \centering
    \includegraphics[width=\linewidth]{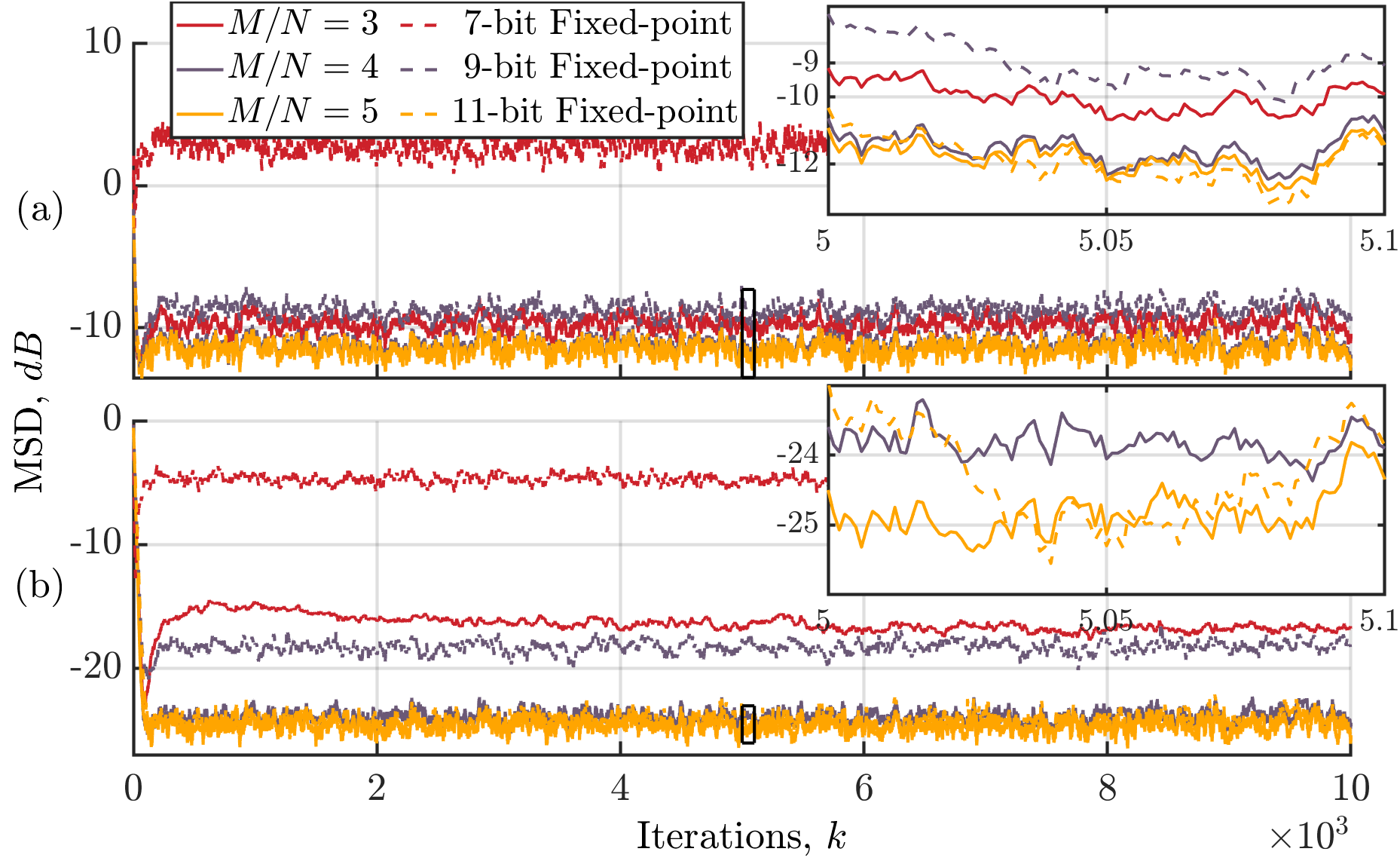}
    \caption{MSD for the SPT-NLMS with different complexities, $M/N$ (SPT terms per coefficient) compared with fixed-point of estimated similar complexity, for (a) $\textrm{SNR}=-3~dB$ and (b) $\textrm{SNR}=10~dB$.}
    \label{fig:NLMS_complexity}
\end{figure}

\begin{figure}[hbt!]
    \centering
    \includegraphics[width=\linewidth]{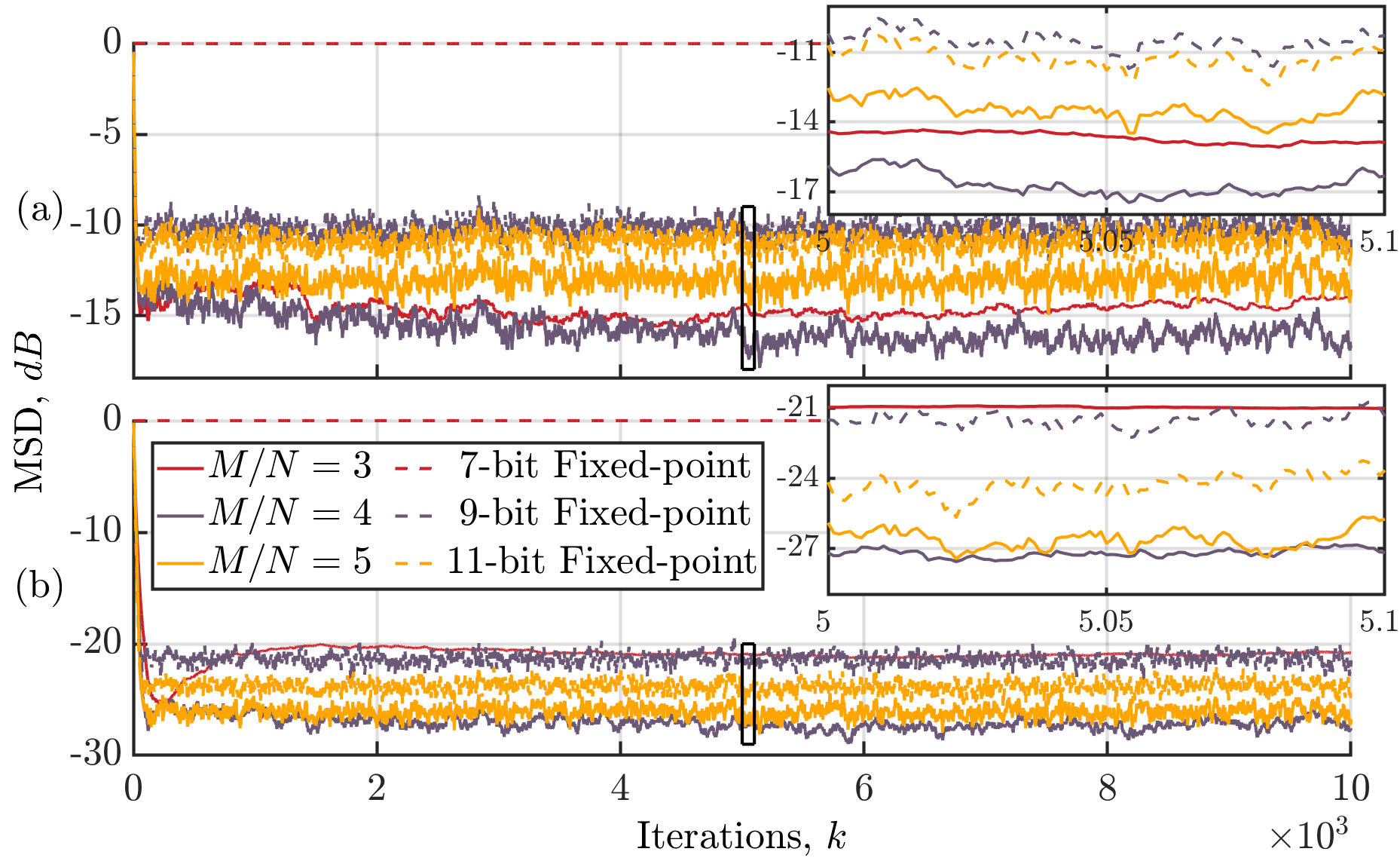}
    \caption{MSD for the SPT-RLS with different complexities, $M/N$ (SPT terms per coefficient) compared with fixed-point of estimated similar complexity, for (a) $\textrm{SNR}=-3~dB$ and (b) $\textrm{SNR}=10~dB$.}
    \label{fig:RLS_complexity}
\end{figure}

Figures~\ref{fig:LMS_complexity},~\ref{fig:NLMS_complexity} and \ref{fig:RLS_complexity} present a comparison of SOPOT filters and its fixed-point counterparts of same estimated complexity. The complexity is estimated considering only the multiplications. As expected, the SPT adaptive filters outperform their fixed-point counterparts. The difference increases at lower complexities, as in the cases involving the 3-SPT filters and the 7-bit fixed-point filters. For the LMS and RLS algorithms, the fixed-point variation cannot adapt itself while the SPT filters have reasonable performance.

\subsection{Sparse System Identification}

In many scenarios, such as channel equalization, echo cancellation and system identification, the unknown systems can present sparse impulse response, containing only few large valued coefficients among many zero (or near zero) valued coefficients~\cite{chen2009sparse,lima2014sparsity}. Thus, we show results for a sparse system identification problem.
In this scenario, depicted in Figure~\ref{fig:SysID}, the unknown system $\bf h$ is a FIR filter with $16$ coefficients, from which only $4$ are non-zero and randomly set. The SNR is set to $30$ dB, and the filter parameter settings are presented in Table~\ref{tab:params2}. The filters are constrained to $M/N=3$ SPT per coefficient, using a $16$-SPT wordlength with the maximum exponent of $3$, i.e., $2^3$ is the highest value an SPT can get.

\begin{table}[ht!]
    \centering
    \caption{Parameter settings for the different adaptive filters used in the identification of the sparse system.}
    \begin{tabular}{|c|c|} \hline
        \textbf{Parameter} & \textbf{SOPOT (Value)} \\ \hline
        $\mu$ & $2^{-5} - 2^{-7} + 2^{-9}$\\ \hline
        $\mu_n$ & $2^{-2} - 2^{-4} + 2^{-6}$ \\ \hline
        $\gamma$ & $2^{-10}$ \\ \hline
        $\lambda$ & $2^0 - 2^{-5} + 2^{-7}$ \\ \hline
        $\delta$ & $2^{-4} - 2^{-9} - 2^{-11}$  \\ \hline
    \end{tabular}
    \label{tab:params2}
\end{table}

\begin{figure}
    \centering
    \includegraphics[width=\linewidth]{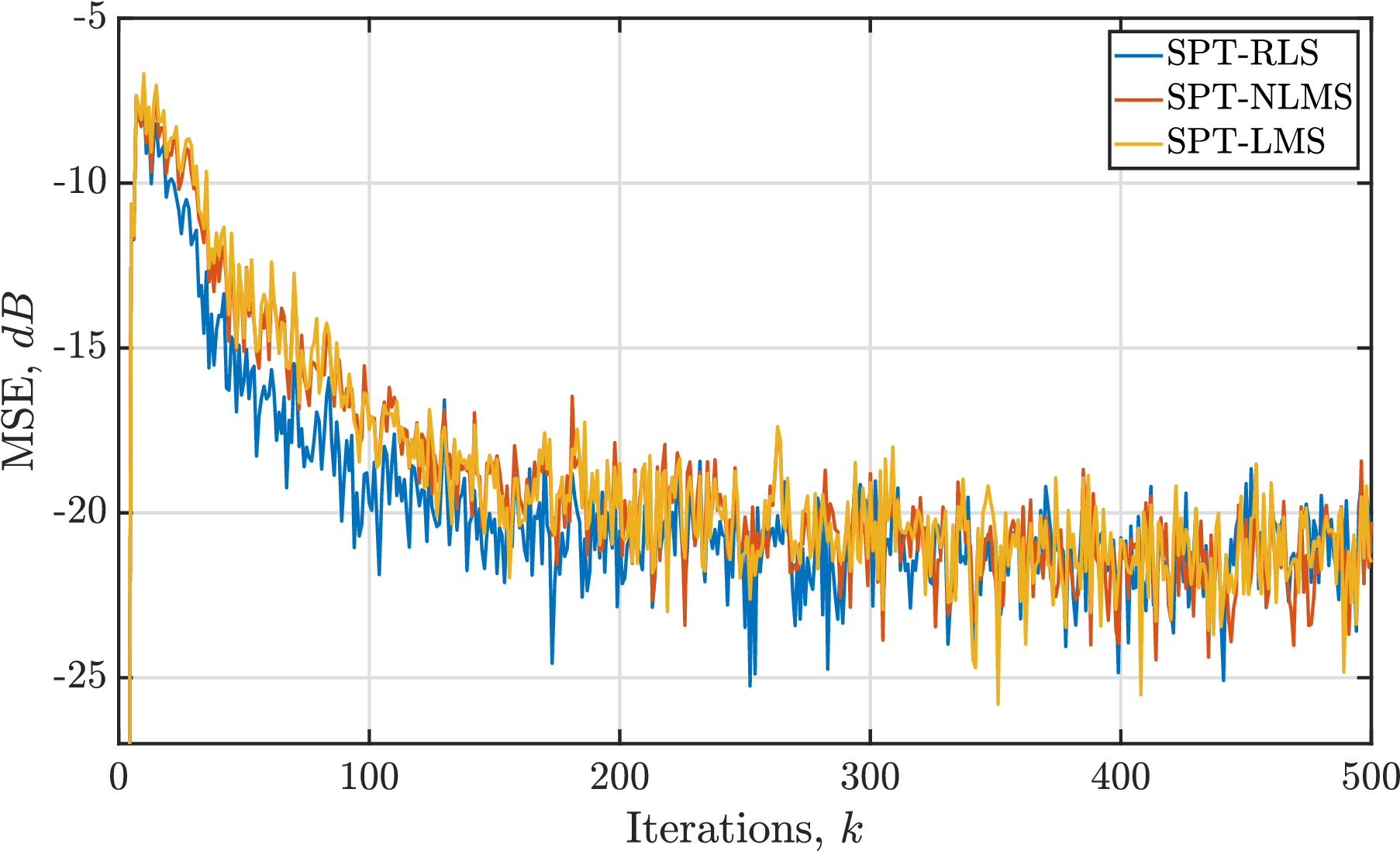}
    \caption{MSE evolution for identifying a sparse unknown system using the SPT-based adaptive filters, the results correspond to a Monte Carlo process of 50 repetitions.}
    \label{fig:sparse_mse}
\end{figure}

\begin{figure}
    \centering
    \includegraphics[width=\linewidth]{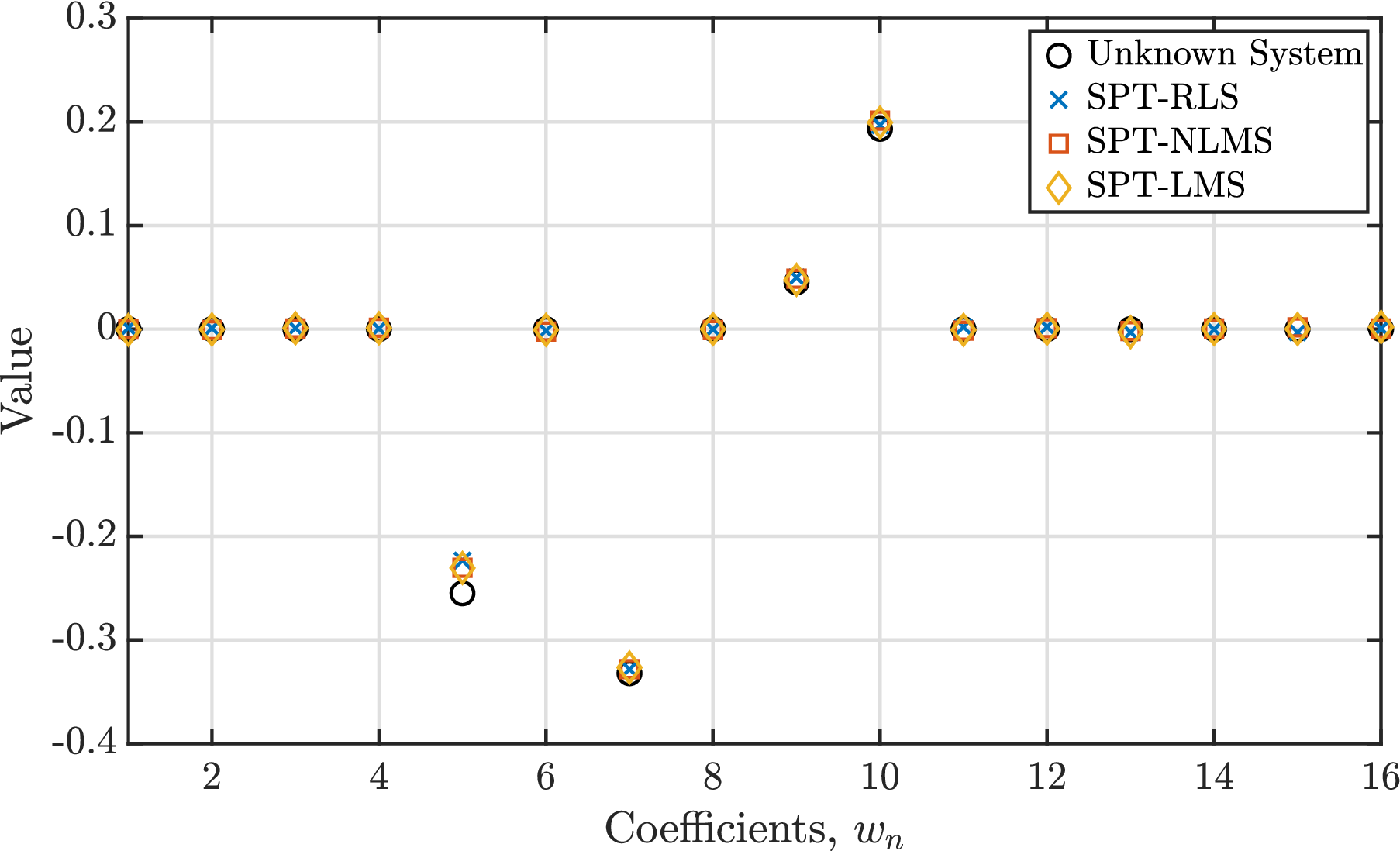}
    \caption{Filter coefficients over the last iteration. The presented values are the average of the coefficients resulting in the last iteration in each of the 50 repetitions of the Monte Carlo process.}
    \label{fig:sparse_coef}
\end{figure}

The results are displayed in Figures~\ref{fig:sparse_mse} and \ref{fig:sparse_coef}, these are averages using $50$ Monte Carlo repetitions. One can observe in Figure~\ref{fig:sparse_mse} the evolution of the MSE through the iterations. The SPT-RLS filter shows slightly faster convergence while there is no visible difference between the SPT-LMS and SPT-NLMS convergence time. In the previous section, similar behavior was observed in the MSD values for the SPT-LMS, SPT-NLMS, and SPT-RLS, with the latter attaining the smaller MSD. The SPT implementation still grants to the RLS relative gains to the LMS and NLMS.

One observes a small difference between the results for the SPT-LMS and SPT-NLMS in Figure~\ref{fig:sparse_mse}. This indicates that the condition number of the input correlation matrix is not large enough to induce a significant difference in their convergence rates~\cite{an1997convergence}. A larger condition number would speed up the convergence of the SPT-RLS algorithm, with the risk of increasing the misalignment in the system identification or, in the worst-case scenario, leading to numerical instability problems due to ill-conditioning~\cite{benesty2005recursive,pierce1992tracking}.

Figure~\ref{fig:sparse_coef} compares the resulting filter coefficients of each SPT-based adaptive filter and the \textit{unknown system}. This phenomenon is accentuated in the fifth element of the coefficient vector. One expects a mismatch in the values of the coefficients due to the high constraint in complexity. It is noticeable that despite that, the SPT-based filters are capable of mimicking the unknown system's impulse response.

\subsection{Change Detection Application}

\begin{figure}[hbt!]
    \centering
    \includegraphics[width=\linewidth]{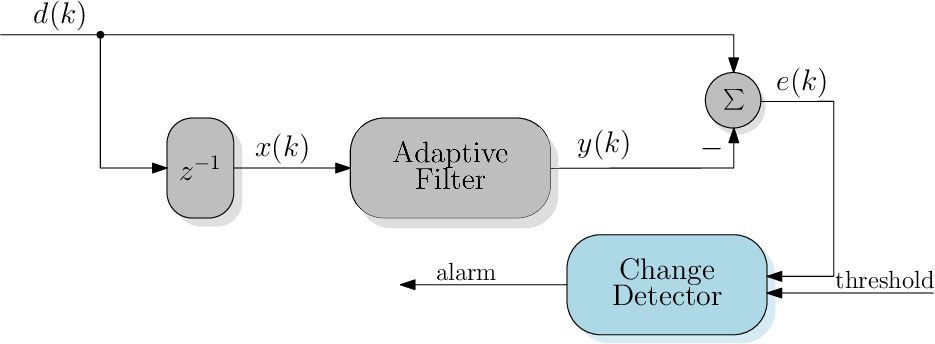}
    \caption{Change detection block diagram: the adaptive filter is used as a predictor, the prediction error signal is then fed to the change detector block.}
    \label{fig:ChangeDetection}
\end{figure}

Change detection is applicable in different engineering areas, ranging from network traffic monitoring~\cite{ahmed2008novel} to fault detection of a system~\cite{gustafsson2000adaptive}. Change detection can be approached using adaptive filters as in Figure~\ref{fig:ChangeDetection}. By using the SOPOT approximations, precision is compromised for the least costly hardware. Nevertheless, this presents possible applications in devices with a limited power source, demanding less energy in calculations. To evaluate possible impacts on change detection applications, one considers adaptive-filters-based spectrum sensing~\cite{e2014spectrum}. One considers the sensing of a band-limited signal in the baseband. The signal originates from a discrete white Gaussian noise signal at 1~Msamples/sec, low-pass filtered using the time-truncated sinc resulting in a limited-band of $142.9~kHz$. This signal is a model for generic baseband signals, such as OFDM signals. It is transmitted through a  Rayleigh fading channel modeled by Jakes' model with a Doppler frequency of $104.2~Hz$. We highlight that this results in a correlated signal. At last, we consider a varying SNR in the detector input.

Following the previous analyses, we compare the performances of fixed point and SOPOT adaptive filters having different complexities. The filters have a short length ($N=4$) and the same complexity (or number density) restrictions as in the previous section. The SOPOT filters employ 3 (low number density), 4 (medium number density), and 5 (high number density) SPTs. The corresponding fixed point word lengths are 7, 9, and 11. Figure \ref{fig:ChangeDetection} shows that for the change detection, it is necessary to fix a threshold. Based on preliminary tests, we test the change detection framework using 27 different values from $2^{-8}+2^{-10}+2^{-12}$  to $2^{-6}-2^{-8}-2^{-12}$ in steps of $2^{-12}$. Therefore, forcing its representation to be within the considered arithmetic restrictions. This experiment does not aim to find the best change detector but instead evaluate the performance compromises using the different implementations.

\begin{figure}[hbt!]
    \centering
    \includegraphics[width=\linewidth]{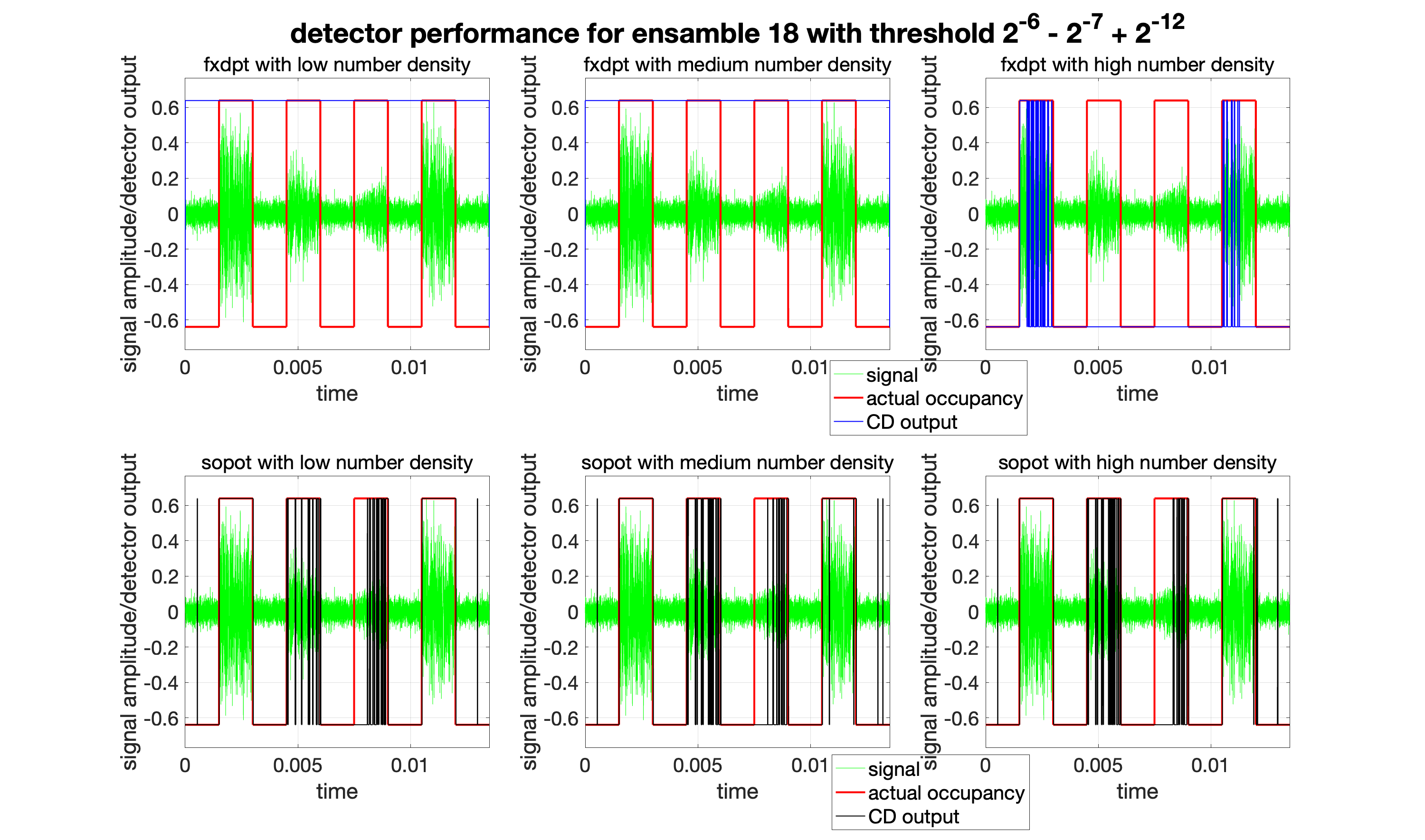}
    \caption{Example of the change detection output using fixed point (top) and SOPOT (bottom) implementations.}
    \label{fig:CD_experiment}
\end{figure}

\begin{figure*}[hbt!]
    \centering
    \includegraphics[width=\linewidth]{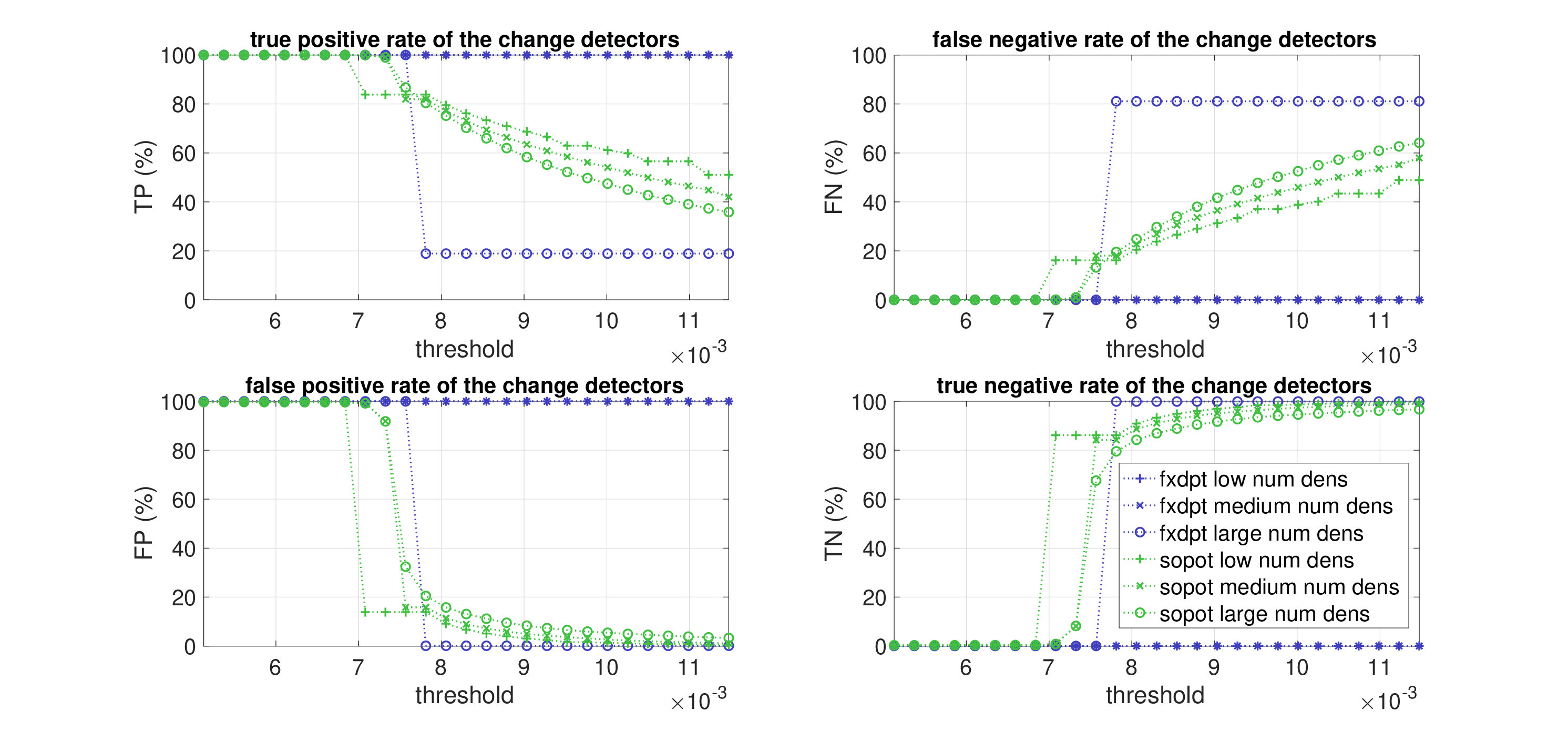}
    \caption{Performance of the change detectors using fixed point and SOPOT implementations.}
    \label{fig:CD_metrics}
\end{figure*}

Figure~\ref{fig:CD_experiment} shows the output for the fixed point and SOPOT change detectors variations under the three complexity constraints. The three graphs on the top correspond to the fixed point implementation and the bottom to the SOPOT implementation. As one notes, the SOPOT implementation delivers an output that is more similar to the actual state of the channel (its occupancy). That is, the SOPOT implementation detects channel occupancy better than the fixed-point implementation does. We apply this procedure to an ensemble of 20 signals (drawn as previously explained, the graphs in Figure~\ref{fig:CD_experiment} correspond to the 18th signal). Figure~\ref{fig:CD_metrics} synthesizes the results in terms of true positive (correct channel occupancy), true negative (correct channel vacancy), false positive (channel vacant and detected as occupied), and false negative (channel occupied and detected as vacant). The values depend on the threshold (horizontal-axis). The performances of the fixed point implementations of low and medium complexities are the same and incapable of providing any useful detection results. On the other hand, the SOPOT implementation provides useful results even at low number density. As the number density increases, the SOPOT implementation has a smoother behavior with the detection threshold.

\section{Conclusion} \label{sec:conclusion}

Low computational complexity adaptive filters are feasible using the SOPOT arithmetic. Despite the SOPOT approximation of the signal in the filter's tapped delay line and the maintenance of the numeric restriction through the calculations, the adaptive filters converge with better results than the fixed-point counterparts under similar computational costs. Exhaustive testing of the filters strengthens this claim. The proposed SPT adaptive filters were tested in different scenarios and presented optimistic results compared with other low-cost alternatives. The numerical issues amplified using fewer bits in the numeric representation did not destabilize the adaptive filtering algorithm.

\ifCLASSOPTIONcaptionsoff
  \newpage
\fi

% \bibliography{thebiblio.bib}
% \bibliographystyle{ieeetr}

\begin{IEEEbiography}[{\includegraphics[width=1in, height=1.25in, clip, keepaspectratio]{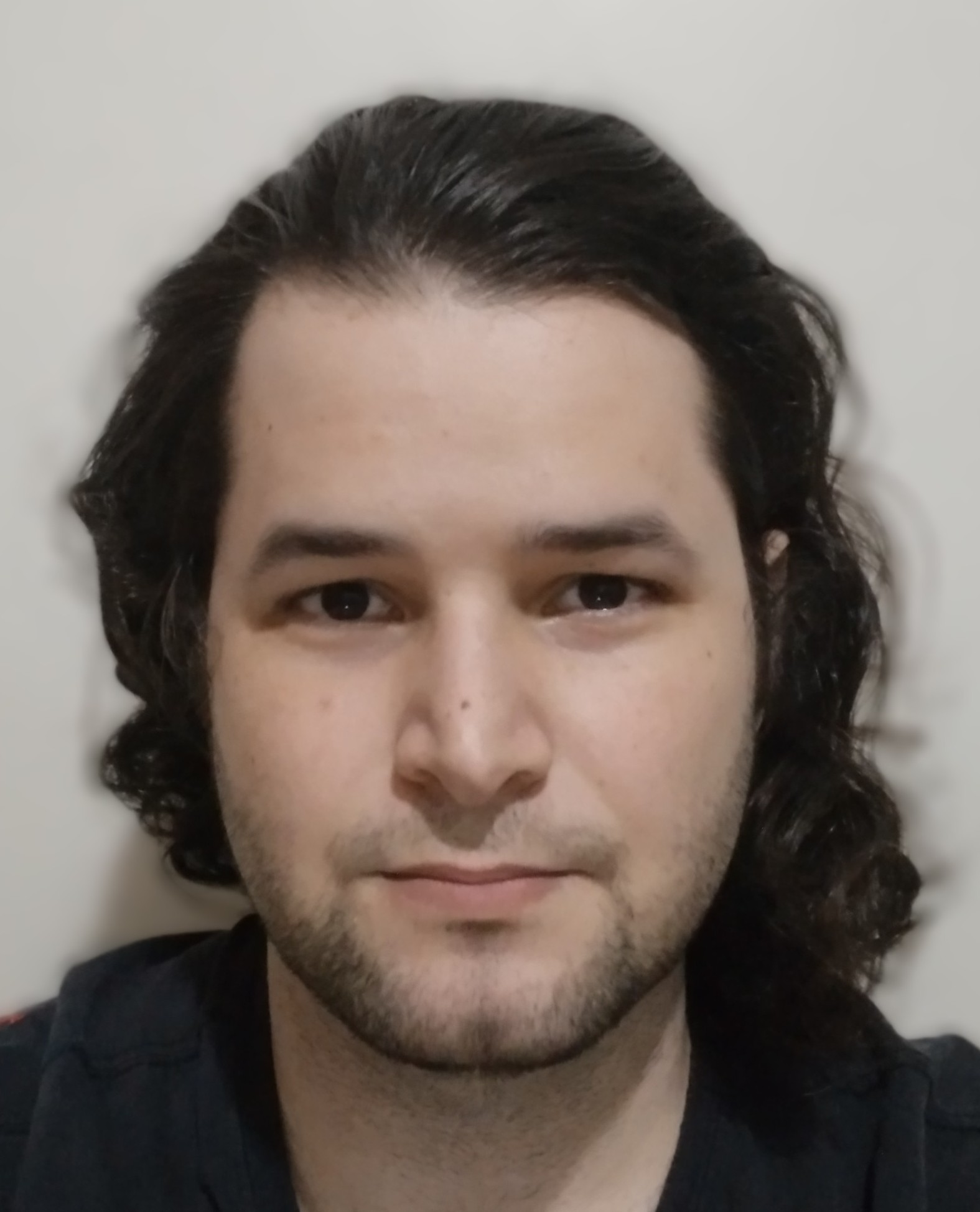}}]{Luiz Felipe Silveira Coelho}
was born in Brasília, Brazil. He obtained the degree in Electrical Engineering with an emphasis in Telecommunications from Rio de Janeiro State University (UERJ) in 2019. Currently (2021) he is pursuing the M.Sc. degree in Electrical Engineering from the Alberto Luiz Coimbra Institute for Graduate Studies and Research in Engineering at the Federal University of Rio de Janeiro (COPPE/UFRJ). His main interests are in the areas of signal processing, communication systems and machine learning applications.
\end{IEEEbiography}

\begin{IEEEbiography}[{\includegraphics[width=1in, height=1.25in, clip, keepaspectratio]{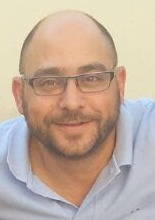}}]{Lisandro Lovisolo}
is an associate professor at the Department of Electronic and Telecommunications Engineering (DETEL) of the State University of Rio de Janeiro (UERJ). At UERJ since 2003, he teaches undergraduate and graduate level courses and has founded PROSAICO (Laboratory of Signal Processing, Intelligent Applications and Communications - www.prosaico.uerj.br) in 2007. His main interests are in the areas of signal processing, image and video, communications systems and networks and the underlying technologies, and computing machinery and its applications. His main interests are in signal processing, wireless communications systems and multimedia communications. He obtained his doctorate in Electrical Engineering from the Federal University of Rio de Janeiro (UFRJ) in 2006. He was born in Neuquen, Argentina in 1974, but feels Brazilian and native from Rio de Janeiro.
\end{IEEEbiography}

\begin{IEEEbiography}[{\includegraphics[width=1in, height=1.25in, clip, keepaspectratio]{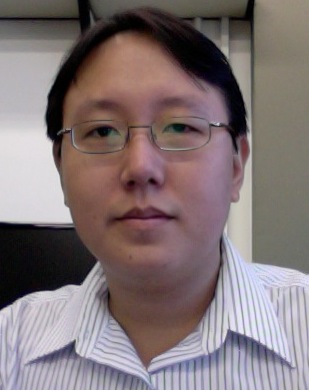}}]{Michel Pompeu Tcheou}
was born in Rio de Janeiro, Brazil. He received the Engineering degree in electronics from Federal University of Rio de Janeiro (UFRJ) in 2003, the M.Sc. and D.Sc. degrees in Electrical Engineering from COPPE/UFRJ in 2005 and 2011, respectively. He has worked at the Electric Power Research Center (Eletrobras Cepel) in Rio de Janeiro, Brazil, from 2006 to 2011. Since 2012 he has been with the Department of Electronics and Telecommunications Engineering at Rio de Janeiro State University (UERJ). He has also been with the Postgraduate Program in Electronics Engineering (PEL). His research interests are in signal processing, communications, data compression, and numerical optimization.
\end{IEEEbiography}

\end{document}